\documentclass[preprint,12pt]{elsarticle}
\usepackage{graphicx}%
\usepackage{multirow}%
\usepackage{amsmath,amssymb,amsfonts}%
\usepackage{amsthm}%
\usepackage{mathrsfs}%
\usepackage[title]{appendix}%
\usepackage{xcolor}%
\usepackage{textcomp}%
\usepackage{manyfoot}%
\usepackage{booktabs}%
\usepackage{algorithm}%
\usepackage{algorithmicx}%
\usepackage{algpseudocode}%
\usepackage{listings}%
\usepackage{graphicx}
\usepackage{booktabs,makecell,multirow}
\usepackage{tabularx}
\usepackage{lipsum}
\usepackage{wrapfig}
\usepackage{algorithm}
\usepackage{algorithmicx}
\usepackage{algpseudocode}

\journal{Neurocomputing}

\begin{document}
\begin{frontmatter}

\title{MCU-Net: A Multi-prior Collaborative Deep Unfolding Network with Gates-controlled Spatial Attention for Accelerated MR Image Reconstruction}

\cortext[cor1]{Corresponding author}
\affiliation[label1]{organization={Chongqing Key Laboratory of Image Cognition, Chongqing University of Posts and Telecommunications},
	city={Chongqing},
	country={China}}
\affiliation[label2]{organization={Key Laboratory of Cyberspace Big Data Intelligent Security, Chongqing University of Posts and Telecommunications, Ministry of Education},
	city={Chongqing},
	country={China}}
\affiliation[label3]{organization={College of Computer and Information Science, Chongqing Normal University},
	city={Chongqing},
	country={China}	
}
\author[label1]{Xiaoyu Qiao}\ead{d210201019@stu.cqupt.edu.cn}
\author[label1,label2]{Weisheng Li\corref{cor1}}\ead{liws@cqupt.edu.cn}
\author[label3]{Guofen Wang}\ead{wanggf@cqnu.edu.cn}
\author[label1]{Yuping Huang}\ead{s180231053@stu.cqupt.edu.cn}

\begin{abstract}
	Deep unfolding networks (DUNs) have demonstrated significant potential in accelerating magnetic resonance imaging (MRI). However, they often encounter high computational costs and slow convergence rates. Besides, they struggle to fully exploit the complementarity when incorporating multiple priors. In this study, we propose a multi-prior collaborative DUN, termed MCU-Net, to address these limitations. Our method features a parallel structure consisting of different optimization-inspired subnetworks based on low-rank and sparsity, respectively. We design a gates-controlled spatial attention module (GSAM), evaluating the relative confidence (RC) and overall confidence (OC) maps for intermediate reconstructions produced by different subnetworks. RC allocates greater weights to the image regions where each subnetwork excels, enabling precise element-wise collaboration. We design correction modules to enhance the effectiveness in regions where both subnetworks exhibit limited performance, as indicated by low OC values, thereby obviating the need for additional branches. The gate units within GSAMs are designed to preserve necessary information across multiple iterations, improving the accuracy of the learned confidence maps and enhancing robustness against accumulated errors. Experimental results on multiple datasets show significant improvements on PSNR and SSIM results with relatively low FLOPs compared to cutting-edge methods. Additionally, the proposed strategy can be conveniently applied to various DUN structures to enhance their performance. 
\end{abstract}


\end{frontmatter}

\section{Introduction}
\label{sec:introduction}
Magnetic resonance imaging (MRI) is an indispensable imaging technique widely employed in clinical scenarios \cite{tu2024novel, chen2024general, yang2023deep}.
However, the complete acquisition of frequency domain (i.e., k-space) raw data is inherently time-consuming, which aggravates patient discomfort and causes image artifacts \cite{harisinghani2019advances}.
To address the problem, sub-Nyquist sampling rates are routinely adopted, resulting in a corresponding reduction in scanning time.
Subsequently, approaches based on parallel imaging (PI) \cite{larkman2007parallel} or compressed sensing (CS) \cite{donoho2006compressed} have been proposed to facilitate the reconstruction of MR images from undersampled data.
PI methods \cite{pruessmann1999sense, griswold2002generalized} exploit the redundancy among k-space data obtained by multiple sensors (receiver coils); however, image quality may deteriorates at high acceleration factors.
In the CS-MRI framework \cite{lustig2007sparse, lustig2008compressed, liu2018cs}, 
prior knowledge such as sparsity \cite{lustig2007sparse} or low-rank \cite{zhao2010low} is typically introduced as regularizers to narrow the solution space. However, CS-based methods still encounter limitations such as reliance on hand-crafted tuning and prolonged iteration processes \cite{hollingsworth2015reducing, qu2010iterative}.

With notable advancements achieved by deep learning (DL) in computer vision tasks \cite{voulodimos2018deep}, various DL-based methods have been proven effective in MRI reconstruction, such as end-to-end training \cite{wang2020deepcomplexmri, akccakaya2019scan}, generative models based on generative adversarial network (GAN) \cite{mardani2018deep, quan2018compressed, noor2024dlgan, li2024progressive, zhou2021efficient} or diffusion models \cite{gungor2023adaptive, cao2024high, wu2023wavelet}, plug-and-play methods \cite{ahmad2020plug, rasti2023plug, kamilov2023plug}, to name a few.
While achieving commendable performance, they also require a significant amount of training data and lack interpretability due to the black-box nature \cite{wang2021deep}.
In comparison, deep unfolding networks (DUNs) \cite{sun2016deep,zhang2020deep,mou2022deep} can be considered as a combination of traditional optimization algorithms and DL techniques.
Various optimization processes are unfolded into deep networks, allowing for the direct learning of parameters and regularizers, which leads to improved reconstruction performance compared to traditional methods.

However, current DUNs generally encounter two significant challenges.
\textit{Firstly}, DUNs face higher computational costs and slower convergence speeds  \cite{ayad2024qn}.
In DUNs, identical blocks, also referred to as cascades \cite{sriram2020end}, are linearly stacked to replicate the iterative steps of optimization algorithms.
While adding more cascades generally improves the performance of DUNs, it also proportionally increases computational costs. This escalation in resource demands can limit the utilization of larger backbone networks in practical applications \cite{heckel2024deep}, necessitating a trade-off.
Moreover, considering the inverse problem is highly ill-posed, errors in shallow cascades can accumulate and be amplified during training, resulting in slower convergence and compromised performance.
Therefore, more efficient design strategy are required in DUNs to guarantee reconstruction performances with a limited number of repetitions.

\textit{Secondly}, when solving the inverse problem, incorporating different prior knowledge into the objective function can help narrow the solution space and better avoid falling into local minima \cite{zhang2023camp, wang2022one, huang2021deep}.
However, the existing DUNs are not customized for multiple prior knowledge and they are inefficient to capture the underlying complementarity.
Some existing methods incorporate multiple priors, but they are simply fused with weighted addition \cite{xu2024attention} or alternative optimization \cite{wang2022one}.
In the case of the former method, the fusing strategy is relatively rudimentary, which may not accurately capture the regions where different optimization-inspired processes excels, thereby necessitating further improvements.
The latter method needs to find a balance from different priors during optimization, which could compromise the convergence speed.

In this study, we propose a multi-prior collaborative DUN called MCU-Net to address the aforementioned problems.
We introduce collaborative cascades (COCAs) designed to construct a novel DUN.
Each COCA comprises three components: optimization-inspired subnetworks (OSNs), a gates-controlled spatial attention module (GSAM), and a correction module (CM). 
First, the OSNs based on sparsity and low-rank produce intermediate reconstructions (IRs) in each COCA.
To fully exploit the underlying complementarity, we design GSAMs to evaluate the relative confidence (RC) and overall confidence (OC) for the IRs.
Considering the underlying correlations of OSNs across COCAs, we design gate units to adaptively preserve necessary long- and short-term information, thereby building robust and precise mappings to the confidence maps.
The IRs are first fused with the RC maps, which assign greater weights to the regions where the corresponding subnetwork excels in an element-wise manner.
Bottleneck scenarios may arise when both subnetworks demonstrate low accuracy in specific locations, as indicated by low OC values, and the weighting process in the GSAMs is unable to effectively enhance the results.
One potential solution to alleviate this issue is the introduction of additional prior knowledge, which can be computationally expensive. 
Instead, we introduce a CM in which the mappings are directly learned from data without incorporating explicit or implicit priors, providing significant flexibility to compensate for errors introduced by the RC-based fusion, obviating the need of introducing extra branches.
When the IRs exhibit sufficiently high OC values, indicating a strong confidence in accepting the RC-fused IRs as the final reconstructions, the correction modules will operate within a relatively small range. 
Otherwise, more weights are assigned to the corrected results to further optimize the fused IRs.
Data consistency (DC) blocks in CMs are used to maintain consistency with the known sampling data.
We adopt floating-point operations (FLOPs) to evaluate the computational costs of different networks.
Experimental results reveal that the proposed method significantly outperforms DUNs employing naive cascades, even with a substantially reduced number of COCAs.
Additionally, when compared to state-of-the-art methods across multiple datasets, our approach achieves superior results while maintaining relatively low FLOPs. 
Furthermore, the proposed method is compatible to optimize the existing DUN based on single or multiple priors.
Experimental results indicate that by implementing the proposed strategy, the performance of various DUNs can be notably enhanced without incurring additional FLOP costs.

The contributions of this study are summarized as follows:
\begin{enumerate}
	\item We propose a novel DUN constructed with collaborative cascades (COCA), for the reconstruction of MR images. The proposed strategy can be conveniently applied to different optimization-inspired network structures to enhance reconstruction performance.
	\item We design a gates-controlled spatial attention module (GSAM) to accurately assess the relative and overall confidence of intermediate results. Building on this, we further develop a two-stage weighting process that facilitates precise element-wise collaboration and tackles bottleneck scenarios for the involved priors.
	\item We conduct extensive experiments across multiple datasets featuring different acceleration rates, sampling masks, and anatomical structures. The results demonstrate that the proposed method robustly achieves superior outcomes with lower computational costs and faster convergence.
\end{enumerate}

\section{Related work}
In this section, we will provide a concise overview of the DUNs applied to MR image reconstruction, based on single or multiple priors.

\subsection{Single-prior-based DUNs}
For MR image reconstruction, a common method is learning the sparse prior in unfolded networks. 
To achieve high-quality reconstructions, different backbones have been adopted.
For example, the authors in \cite{duan2019vs} treated the regularizer as a denoising process, and the mapping was directly learned with cascaded convolutions and ReLUs.
Similarly, the sparsifying transform and its inverse operator in \cite{zhang2018ista} were directly learned with convolutional layers.
Sriram \cite{sriram2020end} et al. extended the variational network \cite{hammernik2018learning} and implemented the update steps in k-space by learning end-to-end, adopting U-Nets as the regularizer.
In ReVarNet \cite{yiasemis2022recurrent}, iterative optimization was performed with convolutional recurrent neural networks (CRNN) in k-space.
In \cite{li2024gates}, the regularizers were designed with convolutions, while ConvLSTM \cite{yu2019review, shi2015convolutional} structure was utilized to filter information across iterations.
Transformer-based networks have also been proposed \cite{huang2022swin, fabian2022humus}, while they may be more computational expensive.
Besides, the regularizer can also be replaced with dual-domain networks \cite{sun2020dual, liu2023diik}, exploiting complementary information from the image and frequency domain.

On the other hand, different algorithms have been utilized to construct the optimization process. In addition to gradient descent \cite{sriram2020end} or proximal gradient descent \cite{hong2024complex}, more complex algorithms have been proven effective, including the alternating direction method of multipliers (ADMM) \cite{sun2016deep, yiasemis2023vsharp}, iterative shrinkage-thresholding algorithm (ISTA) \cite{zhang2018ista}, and half-quadratic splitting (HQS) \cite{jiang2023ga}.
In \cite{wang2024progressive}, the authors designed a dive-and-conquer strategy to decompose the highly ill-posed problem and progressively recover the undersampled regions.
In addition, quasi-Newton methods have been proposed to solve inverse problems, using matrix-valued step size to take more efficient descent steps \cite{ayad2024qn, hong2024complex, ehrhardt2024learning}.

Besides, some methods incorporate low-rank to solve the inverse problem, especially in dynamic MRI \cite{zhang2024t2lr}.
For example, Zhang \cite{zhang2020image} et al. proposed the exploration of simultaneous two-directional low-rankness (STDLR) in k-space data and used an SVD-free algorithm \cite{2004Learning} to reduce computation time.
Some methods adopt structured low-rank (SLR)\cite{jin2016general, jacob2020structured} algorithms to reduce the computational complexity in completing the Hankel matrix.
For example, Pramanik \cite{pramanik2020deep} et al. proposed a general SLR-based network that significantly reduced the computational complexity.
Zhang \cite{zhang2022accelerated} et al. constrained the low rankness of rows and columns of k-space data to reduce computational complexity, and they alleviated the reconstruction errors by introducing prior information.

\subsection{Multi-prior-based DUNs}
To incorporate multiple priors, a common method is introducing different regularization terms in the objective function.
For example, the optimization problem with low-rank and sparse regularization was solved by alternating linearized minimization method in \cite{huang2021deep}.
The authors in \cite{ke2021learned} solved the inverse problem by unfolding ISTA with sparse and low-rank. 
Wang \cite{wang2022one} et al. split a 2D reconstruction problem into several 1D reconstructions and constructed cascaded Hankel matrices on 1D hybrid data.
Sparsity and structured low-rank were incorporated to achieve improved performance.
\cite{xu2024attention} incorporated low-rank prior with image and k-space domain information, while the complementary information was shared with weighted addition operator in the network.
In addition to prior knowledge, some authors proposed to jointly estimate the reconstruction of target images and the optimized sensitivity maps \cite{jun2021joint}, which can also be regarded as a generalized prior. 
However, the existing DUNs are not tailored to fully exploit the complementarity of multiple priors. Our objective is to address this gap.

\section{Problem Formulation}
\label{sec:problem}
In this study, we focus on reconstructing MR images within the framework of PI.
The undersampled $ k $-space measurement $y$ is given as
\begin{equation}
	y= \mathcal{A}x+\epsilon
\end{equation}
where $x$ is the unknown MR image to reconstruct and $\epsilon$ is the measurement noise. 
$\mathcal{A}(\cdot)$ denotes the degradation operator, which is
\begin{equation}
	\mathcal{A}(\cdot)=\mathcal{P}\circ\mathcal{F}\circ\mathcal{E}(\cdot) 
\end{equation}
for PI.
Here, $\mathcal{P}(\cdot)$ denotes the undersampling matrix, $\mathcal{F}(\cdot)$ denotes the Fourier transform and $\mathcal{E}(\cdot)$ is the expanding operator \cite{sriram2020end, yiasemis2022recurrent} given as 
\begin{equation}
	\mathcal E(x)=(\mathcal S_1x,\mathcal S_2x,...,\mathcal S_Cx\big)=(x_1, x_2, ..., x_C)
\end{equation}
that applies sensitivity maps to $x$ to generate coil-specific images.
$C$ is the coil number.
$\mathcal S_i$ denotes the $i$th coil sensitivity map (CSM).
Conversely, the reduced operator is given by
\begin{equation}\label{reduce}
	\mathcal R(x_1,x_2,...,x_C\big)=\sum_{i=1}^C \mathcal S_i^* x_i, i=1, 2, ..., C
\end{equation}
that integrates the coil images into one image. 
Hence, the Hermitian transposition of $\mathcal{A}(\cdot)$ is given by:
\begin{equation}
	\mathcal A^{*}(\cdot)=\mathcal R\circ \mathcal F^{-1}\circ \mathcal P(\cdot)
\end{equation}

To reconstruct MR images from measurement $y$, the objective function in CS-MRI is given as
\begin{equation}\label{sparsity}
	\hat{x}=\mathop{\arg\min}\limits_{x}\|\mathcal A x-y\|_{2}^{2}+\lambda_1 \|\Phi(x)\|_1
\end{equation}
where the first term ensures data fidelity and the second term is a sparsity-based regularizer.
$\lambda$ is a regularization parameter adjusting the ratio.
$\Phi$ is an appropriate transformation field for ensuring the sparsity of the raw data.
This problem can be iteratively solved with proximal gradient descent, and the $k$th iteration is given as
\begin{equation}\label{DC}
	r^k_s=x_s^{k-1}-\alpha^{k}_s\mathcal A^{*}(\mathcal A(x_s^{k-1})-y)
\end{equation}
\begin{equation}\label{DR}
	x_s^k = \mathop{\arg\min}\limits_{x}\|x-r^k_s\|_{2}^{2}+\lambda_1 \|\Phi(x)\|_1
\end{equation}
where $\alpha_s^k$ is the step size.

For a structured low-rank-based method, the objective function is given as
\begin{equation}\label{equ:lr}
	\hat{x}=\mathop{\arg\min}\limits_{x}\|\mathcal A x-y\|_{2}^{2}+\lambda_2 \|\tau(x)\|_*
\end{equation}
where $\tau(\cdot)$ is the lifting operator used to construct hankel matrices. 
Singular value decomposition is routinely used to solve nuclear norms; however, it is computationally intensive and time-consuming.
As an alternative, some studies leveraged the matrix factorization technique \cite{zhang2022accelerated}.
The nuclear norm can be majorized by the iterative reweighted least-squares scheme \cite{mohan2012iterative, wang2022one, pramanik2020deep} as:
\begin{equation}
	\|\tau(x)\|_* \leq \|\tau(x)Q\|_F^2
\end{equation}
and 
\begin{equation}
	Q=[(\tau(x))^H(\tau(x))+\epsilon I]^{-\frac{1}{4}}
\end{equation}
$\|\cdot\|_F^2$ denotes the Frobenius norm.
By leveraging the communicative nature of convolutions, \cite{pramanik2020deep, wang2022one}, we extracted $x^k$ from the lifting operator as
\begin{equation}
	\|\tau(x)Q\|_F^2=\|\mathcal{J}(Q)x\|_F^2
\end{equation}
where $\mathcal{J}(Q)$ is constructed by vertically cascading $\mathcal{D}(q_j)$.
Here, $q_j$ is the $j$-th column of matrix $Q$ and $\mathcal{D}(q_j)$ is a block Hankel matrix constructed from the elements of $q_j$.
Therefore, (\ref{equ:lr}) can be alternatively solved by:
\begin{equation}\label{lowrank}
	x_l^k=\mathop{\arg\min}\limits_{x}\|\mathcal A x-y\|_{2}^{2}+\lambda_2 \|\mathcal{J}(Q^{k-1})x\|_F^2
\end{equation}
\begin{equation}
	Q^k=[(\tau(x^k_l))^H(\tau(x^{k}_l))+\epsilon I]^{-\frac{1}{4}}
\end{equation}
and we solve (\ref{lowrank}) by iterating the following two steps \cite{liu2016projected,zhang2021guaranteed,wang2022one}:
\begin{equation}\label{lr1}
	r_l^k=\lambda_2(\mathcal{J}(Q^{k-1}))^H\mathcal{J}(Q^{k-1})x^{k-1}_l
\end{equation}
\begin{equation}\label{lr2}
	x_l^k=x^{k-1}_l-\alpha^k_l (\mathcal A^{*}(\mathcal A(x^{k-1}_l)-y)+2r_l^k)
\end{equation}

\section{Method}
\label{sec:method}

\begin{figure*}[!t]	
	\begin{minipage}[b]{1\linewidth}
		\centering
		\centerline{\includegraphics[width=13cm]{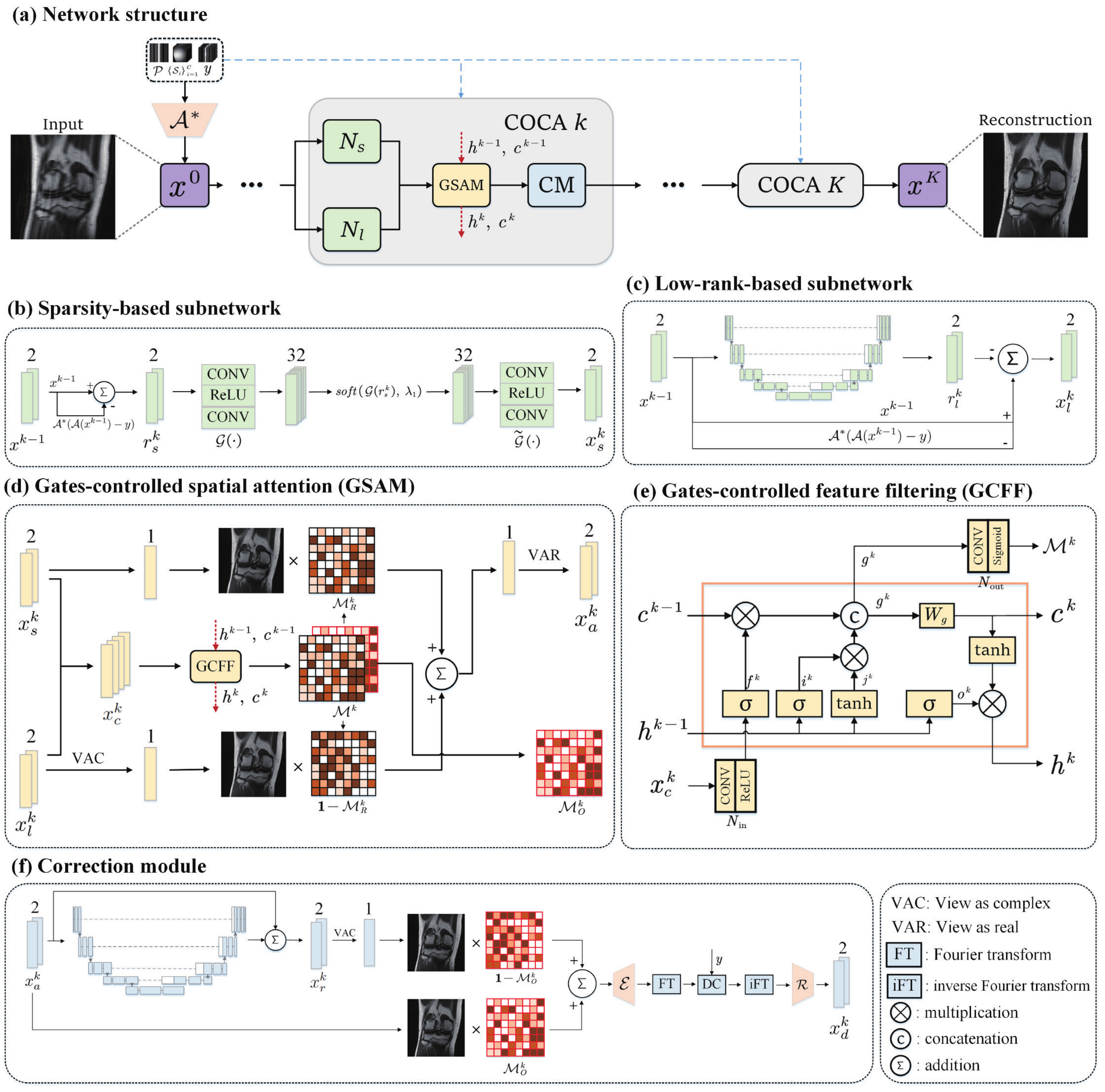}}
	\end{minipage}
	\caption{(a): the structure of MCU-Net. (b): the structure of the sparsity based sub-network ($N_s$ in (a)). (c): the structure of the low-rank based sub-network ($N_l$ in (a)). (d): the structure of GSAMs. (e): the structure of GCFFs in GSAMs. (f) the structure of CMs.
		\label{fig:1}}
\end{figure*}

\subsection{Network Structure}
The network structure of MCU-Net and its components are illustrated in Fig.~\ref{fig:1}.
Fig.~\ref{fig:1} (a) shows the overall workflow.
The undersampled k-space data $y$, the corresponding CSMs $\mathcal\{S_i\}_{i=1}^C$ and the undersampling masks $\mathcal{P}$ are fed to the network.
While there are studies \cite{wang2020deepcomplexmri, fang2024hfgn} based on complex convolutions, it is not the primary focus of our research. In our network, we continue to use two channels to store the real and imaginary data of MR images, enabling a fair comparison with related research \cite{duan2019vs, sriram2020end, yiasemis2022recurrent}.
Besides, the coil-combined image $x^0$ was obtained from the zero-filled measurement $y$ and fed to the network.
Then, $K$ COCAs are stacked to form the main body.
Each COCA comprise three parts: OSNs (green blocks), GSAMs (yellow blocks), and CMs (blue blocks).
Fig.~\ref{fig:1} (b)-(f) depict the detailed structures of the modules, which will be introduced in the following subsections.

\subsection{Optimization-inspired Subnetworks}
Our network features a parallel structure, where we first design sparsity- and low-rank-based subnetworks ($N_s$ and $N_l$ in Fig.~\ref{fig:1} (a)) in each COCA, aiming at capturing complementary information to improve reconstruction performance. 
For $N_s$, we adopt ISTA-Net \cite{zhang2018ista} as the backbone network, where the sparse transform in (\ref{DR}) is directly learned during training, given as:
\begin{equation}\label{ISTA-learned}
	x_s^k=\mathcal{\widetilde{G}}(soft(\mathcal G(r_s^k), \lambda_1))
\end{equation}
where $\mathcal G(\cdot)$ is the learned transform and $\mathcal{\widetilde{G}}(\cdot)$ is its inverse operator constrained by a customized loss function. $\lambda_1$ is the shrinkage threshold. 
$\mathcal G(\cdot)$ and $\mathcal{\widetilde{G}(\cdot)}$ are implemented using cascaded convolutional layers separated by activation functions, as shown in Fig.~\ref{fig:1} (b).
Therefore, $N_s$ denotes the iterative process in (\ref{DC}) and (\ref{ISTA-learned}), and we have
\begin{equation}\label{eq:xsk}
	x^k_s = N_s(x^{k-1})
\end{equation}

The structure of $N_l$ in each COCA is depicted in Fig.~\ref{fig:1} (c), which is constructed by unrolling (\ref{lr1}) and (\ref{lr2}) into the network. 
For (\ref{lr1}), $(\mathcal{J}(Q))^H\mathcal{J}(Q)x$ can be regarded as a projection acting on $x$ which is determined by $Q$, and $Q$ at each iteration will also be updated according to the status of $x$.
Therefore, we relax the constraint on (\ref{lr1}) and we adopt a U-Net, labeled as $N_{lu}$, to learn the mapping directly, given as
\begin{equation}\label{lr_unet}
	r_l^k = \lambda_2 N_{lu}(x^{k-1})
\end{equation}
Similar as (\ref{eq:xsk}), we obtain
\begin{equation}
	x^k_l = N_l(x^{k-1})
\end{equation}
where $N_l$ denotes the processes of (\ref{lr_unet}) and (\ref{lr2}).
With different underlying driving algorithms and optimizing-inspired structures, the IRs from different subnetworks show varying reconstruction performances, which offer substantial complementarity.

\subsection{Gates-controlled Spatial Attention}
To fully leverage the complementary information from different optimization-inspired process, we design a GSAM to identify the regions where different OSNs excels, whose structure is detailed in Fig.~\ref{fig:1} (d).
To achieve this, we first design RC maps to precisely weight the IRs produced by different OSNs in an element-wise manner. 
The RC essentially indicates the confidence level to accept each pixel value as the final predictions. 
Consequently, when the predicted value from one subnetwork matches the target closer than the other one, the RC value should be closer to one, signifying high confidence.
We use $\mathcal{M}^k_R$ to represent the RC map in the $k$-th COCA, and during the weighting process, we transfer $x_s^k$ and $x_l^k$ back to complex values and then implement a Hadamard product between $\mathcal{M}^k_R$ and $x_s^k$. 
$x_l^k$ is similarly multiplied by $\textbf{1}-\mathcal{M}^k_R$. 
We use $N_a$ to represent the weighting process, and the weighted output is given by
\begin{equation}
	x^k_a = N_a(x_s^k, x_l^k; \mathcal{M}^k_R) = x^k_s \odot \mathcal{M}^k_R + x^k_l \odot (\textbf{1}-\mathcal{M}^k_R)
\end{equation}
where $x_s^k$ and $x_l^k$ are IRs produced by subnetworks in the $k$-th COCA, $x^k_a$ is the RC-weighted result and $\textbf{1}\in\mathbb{R}^N$ is a matrix filled with one. $\odot$ denotes the Hadamard product operator.

On the other hand, there could be bottleneck scenarios where both OSNs show limited performances.
To address the issue, we introduced OC maps, which indicate the confidence level of $x^k_a$, while CMs are designed to further optimize the low-OC areas.
Details of the CMs will be elaborate in section \ref{CM}.
Consequently, the GSAMs are designed to simultaneously learn the RC and OC of the IRs in different COCAs.

Empirically, there should be underlying correlations among the regions where different OSNs excels across COCAs.
Therefore, integrating long- and short-term information from different COCAs may potentially facilitate the construction of stable mappings to the target confidence maps.
Inspired by the long short-term memory (LSTM) technique and the variations based on convolutions \cite{shi2015convolutional}, we design gates-controlled feature filtering (GCFF) module in each GSAM to adaptively filter necessary information from all the preceding COCAs.
The structure of GCFF is depicted in Fig.~\ref{fig:1} (e), and the computational process can be expressed as:
\begin{equation}\label{LSTM}
	\begin{aligned}
		&x_c^k = \mathrm{concat}(x_s^k, x_l^k)  \\
		&x_{\mathrm{in}}^k = N_{\mathrm{in}}(x_c^k) \\
		&f^k = \sigma(W_{xf} \ast x_{\mathrm{in}}^k + W_{hf} \ast h^{k-1} +b_f)\\		
		&i^k= \sigma(W_{xi} \ast x_{\mathrm{in}}^k + W_{hi} \ast h^{k-1} +b_i)\\
		&j^k = \mathrm{tanh}(W_{xj} \ast x_{\mathrm{in}}^k + W_{hj} \ast h^{k-1} +b_j)\\
		&g^k = \mathrm{concat}(f^k \circ c^{k-1}, i^k \circ j^k)\\
		&\mathcal{M}^k = N_{\mathrm{out}}(g^k)\\
		&c^k = W_g \ast g^k\\
		&o^k = \sigma(W_{xo} \ast x_{\mathrm{in}}^k + W_{ho} \ast h^{k-1} +b_o)\\
		&h^k = o^k \circ \mathrm{tanh}(c^k)
	\end{aligned}
\end{equation}
where $\sigma$ denotes Sigmoid functions, $W$ denotes convolutions and $b$ is the bias.
First, the IRs from current OSNs are concatenated and lifted to the feature space with a convolution and a ReLU function ($N_{\mathrm{in}}$), obtaining $x_{\mathrm{in}}^k$.
The forget gate $f^k$ discards irrelevant information from the cell state $c^{k-1}$, while the input gate modulate new information introduced from $x_{\mathrm{in}}^k$.
These information are concatenated as $g^k$, which involves necessary information from the current and previous COCAS, and we utilize a convolution with a Sigmoid function ($N_{\mathrm{out}}$) to learn the confidence maps $\mathcal{M}^k$ from $g^k$.
After that, a $1\times1$ convolution $W_g$ is adopted to shrinkage the channel number and update the cell state.
Finally, the new hidden state is modulated with the new cell state $c^k$ and the output gate ($o^k$).

The confidence map $\mathcal{M}^k$ learned in OSNs has two channels, given as:
\begin{equation}
	c^k, h^k, \mathcal{M}^k = N_{\mathrm{GCFF}}(x^k_s, x^k_l, c^{k-1}, h^{k-1})
\end{equation}
where $N_{\mathrm{GCFF}}$ denotes the process in (\ref{LSTM}). Each element in $M^k$ lies between zero and one.
We separate $\mathcal{M}^k$ into $\mathcal{M}^k_R$ and $\mathcal{M}^k_O$, denoting RC and OC maps, respectively.
The OC maps as well as the weighted IRs (\textit{i.e.}, $x^k_a$) are fed to the CMs in the current COCA, while the updated cell and hidden state ($h^k$ and $c^k$) are fed to the GCFF in the next COCA.

\subsection{Correction Modules}\label{CM}
Through GSAMs, the pixels reconstructed by proper subnetworks are assigned with larger weights, and the gap between intermediate outputs and targets can be narrowed.
However, considering that we only experiments the collaboration between two OSNs, there may be regions where both OSNs show limited performance. For example, the intermediate values from both subnetworks are significantly bigger or smaller than the target values.
In such cases, the RC-based weighting process encounters challenges in bringing the intermediate reconstructions closer to the target.
Incorporating additional priors may alleviate the issue; however, the associated computational costs are significant.
Besides, the weighting process in GSAMs may also introduce new errors, leading to suboptimal convergence.
Therefore, we propose a CM address these problems.

In each CM, a correcting network $N_r$ is first utilized to improve the quality of $x_a^k$ as:
\begin{equation}\label{regunet}
	x_r^k = x_a^k + N_r(x_a^k)
\end{equation}
where $N_r$ is implemented by a U-Net backbone.
We design a residual structure to make $N_r$ directly learn the updated information.
Unlike subnetworks, conventional prior knowledge is not explicitly or implicitly involved in narrowing the solution space, and $N_r$ directly learn the mapping from data.
When the outputs from GSAMs are of sufficient accuracy for some pixels, replacing them with the corrected results could compromise the performance.
Therefore, we adopt the OC maps produced in GSAMs, which represent the confidence to accept the GSAM-based IRs as the final results in an element-wise manner.
The OC-based correction process is given as:
\begin{equation}\label{oc}
	\begin{aligned}
		x_o^k &= N_o(x_a^k, x_r^k; \mathcal{M}^k_O)\\
		&= x_a^k \odot \mathcal{M}^k_O + x_r^k \odot (\textbf{1}-\mathcal{M}^k_O) \\
		&= x_a^k \odot \mathcal{M}^k_O + (x_a^k + N_r(x_a^k))\odot (\textbf{1}-\mathcal{M}^k_O)
	\end{aligned}
\end{equation}
where $x_o^k$ is the OC-weighted output. 

Finally, we utilize a data consistency (DC) block after the OC-based weighting to update the $m$th coefficient in the reconstructed images to keep consistency with known measurements as:
\begin{equation}\label{dc}
	y^k_d(m) =\left\{ 
	\begin{array}{l}
		y^k_o(m), \ \ \ \ \ \ \ \ \ m\notin\Omega\\
		\frac{y^k_o(m)+\mu^k y(m)}{1+\mu^k}, \ \ \ \ m\in\Omega\\
	\end{array} \right.
\end{equation}
where $y^k_d$ is the output of the DC blocks, $y^k_o$ is the multi-coil k-space data of $x^k_o$ and $\Omega$ is a subset of fully-sampled k-space data, representing sampled data points.
For the sampled point, a learned $\mu^k$ is used to adjust the ratio between the measurements and the reconstructed values.
Finally, $y^k_d$ is transferred back to the image domain using inverse Fourier transform (iFT) and the reduce operator, and the optimized $x^k_d$, which is equal to $x^{k}$, is fed into the $(k+1)$th COCA.
Our approach enhances the reconstruction results in each COCA, enabling the subsequent COCA to have a better chance of approaching the target images, thus improving the overall network performance.

The complete workflow of MCU-Net is given in Algorithm \ref{alg1}.

\begin{algorithm} 
	\caption{MCU-Net} 
	\label{alg1} 
	\begin{algorithmic}[1]
		\renewcommand{\algorithmicrequire}{ \textbf{Input: }}
		\Require Mask $\mathcal{P}$, CSMs $\mathcal\{S_i\}_{i=1}^C$, Measurement $y$
		\renewcommand{\algorithmicrequire}{ \textbf{Set: }}
		\Require $K=10$
		\renewcommand{\algorithmicrequire}{ \textbf{Output: }}
		\Require Reconstructed image $x^K$
		\State $x_0 \gets \mathcal{A}^*y$
		\State $c_0 \gets W_\mathrm{Init} (x_0)$
		\For{$k\in\{1, 2, ..., K\}$}
		{
		\State $x^k_s, x^k_l \gets N_s(x^{k-1}), N_l(x^{k-1})$
		\If{$k=1$}
		\State $c^k, h^k, \mathcal{M}^k \gets N_{\mathrm{GCFF}}(x^k_s, x^k_l, c^{k-1})$
		\Else
		\State $c^k, h^k, \mathcal{M}^k \gets N_{\mathrm{GCFF}}(x^k_s, x^k_l, c^{k-1}, h^{k-1})$
		\EndIf
		\State $\mathcal{M}^k_R, \mathcal{M}^k_O \gets \mathcal{M}^k $
		\State $x^k_a \gets x^k_s \odot \mathcal{M}^k_R + x^k_l \odot (\textbf{1}-\mathcal{M}^k_R)$
		\State $x_o^k \gets x_a^k \odot \mathcal{M}^k_O + (x_a^k + N_r(x_a^k))\odot (\textbf{1}-\mathcal{M}^k_O) $
		\State $x^k \gets \mathrm{DC}(x_o^k)$
		\EndFor
		}
	\end{algorithmic}
\end{algorithm}

\subsection{Settings and Loss Function}
We set the number of COCAs ($K$) to 10 in the network.
$\alpha^k_s$, $\alpha^k_l$, $\lambda_1$, $\lambda_2$, $\mu^k$ were directly learned in the network.
For $N_s$, the number of channels in the convolutional layers was set to 32, 32 in $\mathcal{G}(\cdot)$, and 32, 2 in $\mathcal{\widetilde{G}(\cdot)}$. The kernel size in the convolutional layers was set to 3 and the padding size was set to 1.
For U-Net in $N_l$, the number of pooling layers was set to four and the number of channels was set to 16, 32, 64, and 128.
For the convolutions in GSAMs, the hidden channel number was set to 8.
The kernel size in $W_g$ was set to 1, and it was set to 3 in the other convolutions. 
For U-Net in the correction modules, the number of pooling layers was set to 4, and the number of channels was set to 4, 8, 16, and 32. The initial cell state $c_0$ is learned from $x_0$ with a convolution $W_{\mathrm{init}}$.

The experiments were conducted on a RTX 3090 GPU (24GB).
During training, the batch size was set to 1, and the learning rate was set to 1e-3.
The network was trained by a composited loss given as
\begin{equation}
	\begin{aligned}
		\mathcal{L} = \sum_{k=1}^{K}[\gamma_1^k\mathcal{L}_{\mathrm{MSE}}(x_g, x^k)
		+ \gamma_1^k\mathcal{L}_{\mathrm{SSIM}}(x_g, x^k) + \gamma_2\| \mathcal{\widetilde{G}}(\mathcal{G}(r_s^{k}))-r_s^{k}\|^2_2]
	\end{aligned}
\end{equation}
where the final output $x_K$ and the intermediate outputs from each COCA are fed to the loss function to supervise the corresponding modules.
$x_g$ is the target image (ground truth) and $K$ is the total number of stacked COCAs. $\mathcal{L}_{\mathrm{MSE}}$ and $\mathcal{L}_{\mathrm{SSIM}}$ denotes mean square error (MSE) loss and structural similarity index (SSIM) loss \cite{zhao2015loss}, respectively.
In this study, $\{\gamma_1\}_{k=1}^K$ is an increasing sequence to assign different weights to the outputs from different COCAs \cite{yiasemis2023vsharp} and $\gamma_1^k=10^{\frac{k-K}{K-1}}$
The last term was used to fulfill the symmetry constraint of $\mathcal{G}$ and $\widetilde{\mathcal{G}}$, and $\gamma_2$ was set to 0.01.
In addition, when evaluating on the coronal PD sequence, we removed SSIM loss since MCU-Net achieve satisfactory results with MSE loss.

\section{Experimental Results}
\label{results}

In this section, we compared the reconstruction performance of MCU-Net with cutting-edge methods on various datasets to demonstrate its superiority. Ablation studies were designed to highlight the necessity of the proposed modules. Furthermore, we extended the collaborative strategy to various DUNs, showcasing its versatility and effectiveness with different optimization-inspired processes.

\subsection{Datasets}
\begin{figure}[!t]	
	\begin{minipage}[b]{1\linewidth}
		\centering
		\centerline{\includegraphics[width=8.5cm]{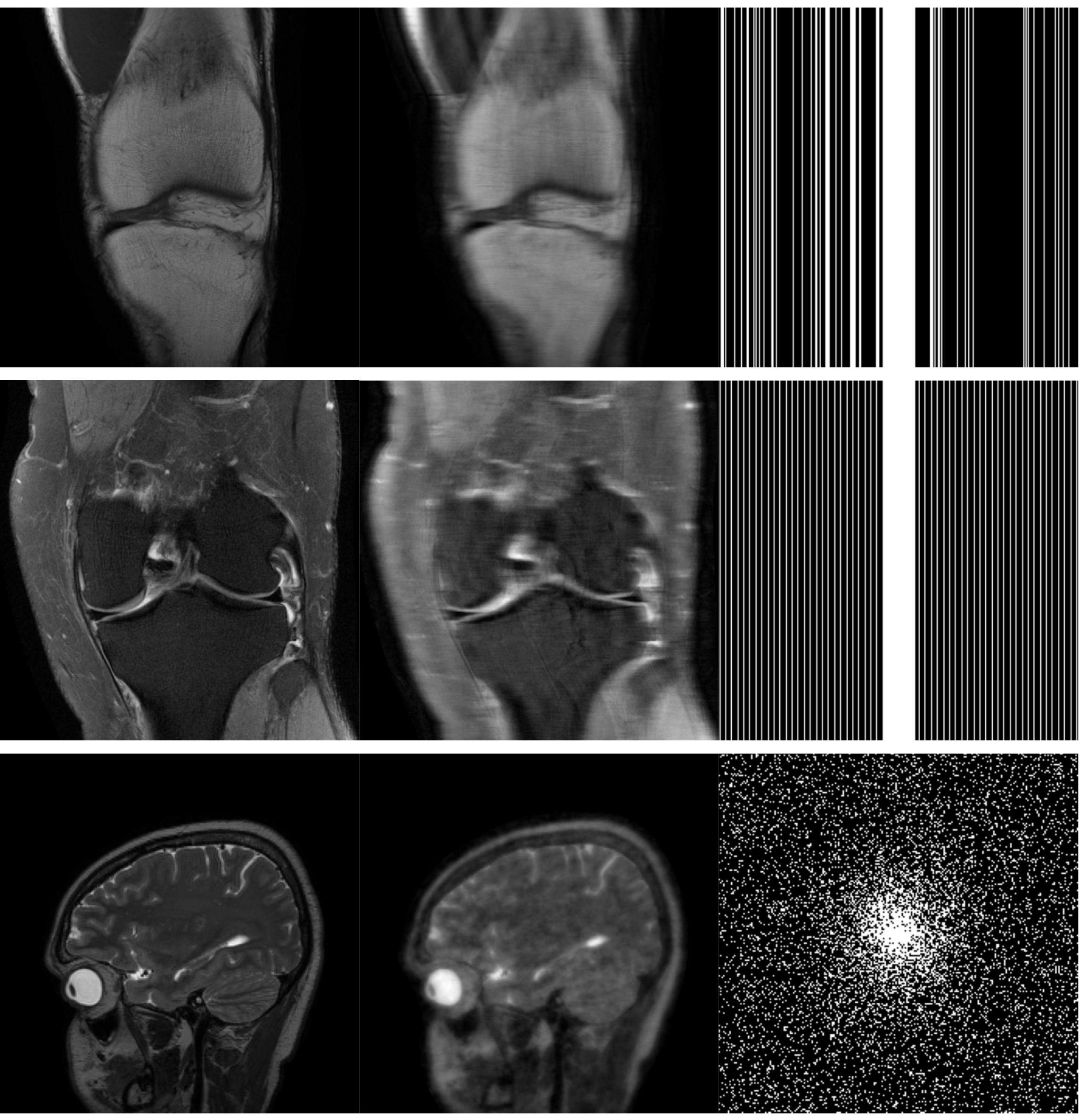}}
	\end{minipage}
	\caption{Examples of training data in the NYU dataset (the first row), FastMRI dataset (the second row) and the brain dataset (the third row). The first column displays the ground truth images, the second column shows the zero-filled images and the third column are the corresponding sampling masks.
		\label{fig:2}}
\end{figure}

In this study, we utilized three datasets to evaluate our methods.
The first dataset is a publicly available NYU knee dataset \cite{pramanik2020deep, hammernik2018learning, duan2019vs} comprises five sequences:
coronal proton density (PD), coronal fat-saturated PD (PDFS), axial fat-saturated T2, Sagittal fat-saturated T2 and Sagittal PD.
Each sequence contains MR images from 20 patients, with each patient comprising approximately 40 slices.
In our experiments, The training, validation, and testing data are divided in an approximate ratio of 3:1:1. None of the validation or testing data was used for training, and vice versa. 

We also conducted our experiments utilizing the FastMRI dataset \cite{zbontar2018fastmri}. Due to the limitations of our training hardware, we trained the networks on a subset of the FastMRI knee dataset.
Specifically, 1199 slices were used to train the networks with the image size of 640 $\times$ 368.
Because the FastMRI dataset did not provide fully sampled test data, we partitioned the validation set into a test set and a new validation set. 
During training, 100 slices were used for validation.
Finally, 950 slices were used to evaluate the network performance.
All the data were randomly selected, ensuring no overlap among them.

In addition to these two knee datasets, we incorporated a brain dataset \cite{aggarwal2018modl} to further strengthen our findings. 
This dataset contains 360 slices for training and 164 slices for testing with the image size of 256 $\times$ 232. 
We randomly selected 10 slices from the test set for validation, leaving 154 slices for evaluating the network performance.

For the NYU dataset and FastMRI datast, we applied random and equispaced Cartesian undersampling with 4-fold and 6-fold acceleration, respectively, sampling 24 lines at the central region.
For the NYU datasets, the CSMs were pre-computed using ESPIRiT \cite{uecker2014espirit}, and they are provided in the dataset.
For FastMRI dataset, we pre-computed the corresponding CSMs using the Sigpy \cite{ong2019sigpy} toolbox.
For the brain dataset, CSMs and 2D random sampling masks are also included in the dataset.
Examples of MR images and the corresponding sampling masks are shown in Fig.~\ref{fig:2}.
More detailed data collection parameters of the aforementioned dataset are available in the corresponding articles \cite{hammernik2018learning, zbontar2018fastmri, aggarwal2018modl}.

\subsection{Comparison with cutting-edge methods}
\begin{table*}[!t]
	\centering
	\caption{The $4\times$ and $6\times$ quantitative results on the NYU and FastMRI dataset compared with the state-of-the-art methods.}\label{sota}
	\resizebox{1.\columnwidth}{!}{
		\begin{tabular}{llllll} 
			\toprule
			\multirow{2}{*}{Sequence} & \multirow{2}{*}{Method} & \multicolumn{2}{l}{PSNR$\uparrow$} & \multicolumn{2}{l}{SSIM$\uparrow$} \\\cline{3-6}
			& & $4\times$ & $6\times$ & $4\times$ & $6\times$ \\
			\hline 
			\multirow{10}{*}{Coronal PD} & zero-filled & 31.3089$\pm$3.3939 & 30.5176$\pm$3.3902 & 0.8778$\pm$0.0713 & 0.8626$\pm$0.0799\\
			& TV & 33.7498$\pm$3.1753 & 32.3290$\pm$3.2159 & 0.8851$\pm$0.0657 & 0.8658$\pm$0.0725\\
			& U-Net & 36.8984$\pm$2.7072 & 34.7352$\pm$2.7863 & 0.9361$\pm$0.0610 & 0.9133$\pm$0.0682\\
			& D5C5  & 38.7903$\pm$2.8408 & 35.9697$\pm$2.7990 & 0.9509$\pm$0.0587 & 0.9259$\pm$0.0689 \\
			& ISTA-Net & 39.2910$\pm$2.8423 & 35.5220$\pm$3.0915 & 0.9471$\pm$0.0567 & 0.9138$\pm$0.0717\\
			& VS-Net & 40.1649$\pm$2.8767 & 37.0604$\pm$2.7927 & 0.9591$\pm$0.0569 & 0.9358$\pm$0.0648\\
			& E2E-VarNet & 39.3592$\pm$2.9368 & 36.8949$\pm$3.3872 & 0.9542$\pm$0.0662 & 0.9342$\pm$0.0775\\
			& ReVarNet & 40.2545$\pm$2.8766 & 37.3683$\pm$2.7869 & 0.9590$\pm$0.0588 & 0.9379$\pm$0.0677\\
			& MeDL-Net & 40.9295$\pm$2.9072 & 38.1052$\pm$2.7622 & 0.9614$\pm$0.0557 & 0.9417$\pm$0.0663\\
			& vSharp & 40.3121$\pm$2.7773 & 37.5657$\pm$2.5346 & 0.9341$\pm$0.0575 & 0.8982$\pm$0.0660\\
			& MCU-Net (ours) & \textbf{41.2003$\pm$2.9637} & \textbf{38.4734$\pm$2.8095} & \textbf{0.9641$\pm$0.0595} & \textbf{0.9455$\pm$0.0668}\\
			\hline 
			\multirowcell{10}{Coronal PDFS} & zero-filled & 32.0299$\pm$2.0273 & 31.2221$\pm$2.0575 & 0.7964$\pm$0.0998 & 0.7587$\pm$0.1172\\
			& TV & 33.7256$\pm$1.9058 & 32.7328$\pm$1.8850 & 0.8124$\pm$0.0752 & 0.7920$\pm$0.0796\\
			& U-Net & 34.4367$\pm$0.6572 & 33.2028$\pm$2.5662 & 0.8183$\pm$0.0106 & 0.7758$\pm$0.1222\\
			& D5C5& 34.9671$\pm$2.8261 & 33.4189$\pm$2.4765 & 0.8250$\pm$0.1106 & 0.7854$\pm$0.1223 \\
			& ISTA-Net & 34.4990$\pm$2.6389 & 33.4569$\pm$2.5758 & 0.8198$\pm$0.1066 & 0.7854$\pm$0.1214\\
			& VS-Net & 34.6847$\pm$2.7134 & 33.5010$\pm$2.6100 & 0.8259$\pm$0.1073 & 0.7842$\pm$0.1229\\
			& E2E-VarNet & 34.6274$\pm$2.7331 & 33.1626$\pm$2.4742 & 0.8234$\pm$0.1101 & 0.7804$\pm$0.1244 \\
			& ReVarNet & 34.8522$\pm$2.7461 & 33.3146$\pm$2.5604 & 0.8272$\pm$0.1080 & 0.7832$\pm$0.1232\\
			& MeDL-Net & 35.5931$\pm$2.7307 & 34.1409$\pm$2.5501 & 0.8349$\pm$0.1025 & 0.7922$\pm$0.1255\\
			& vSharp & 35.5536$\pm$2.6244 & 34.3442$\pm$2.4485 & 0.8495$\pm$0.0940 & 0.8210$\pm$0.0990\\
			& MCU-Net (ours) & \textbf{36.1873$\pm$2.5950} & \textbf{34.7850$\pm$2.3066} & \textbf{0.8685$\pm$0.0890} & \textbf{0.8468$\pm$0.0917}\\		
			\hline 
			\multirow{10}{*}{FastMRI} & zero-filled & 32.8883$\pm$3.0920 & 32.0634$\pm$3.0960 & 0.8856$\pm$0.0681 & 0.8679$\pm$0.0758\\
			& TV & 35.3960$\pm$2.8296 & 33.6309$\pm$2.7932 & 0.8920$\pm$0.0610 & 0.8696$\pm$0.0704\\
			& U-Net & 36.5615$\pm$3.0155 & 35.2409$\pm$2.9859 & 0.9188$\pm$0.0622 & 0.8959$\pm$0.0694\\
			& D5C5 & 37.4408$\pm$3.3691 &35.6572$\pm$3.0195  & 0.9297$\pm$0.0638 & 0.9056$\pm$0.0712\\
			& ISTA-Net & 36.8610$\pm$3.3105 & 34.9415$\pm$3.2019 & 0.9260$\pm$0.0655 & 0.8917$\pm$0.0698\\
			& VS-Net & 37.6398$\pm$3.4316 & 35.0945$\pm$3.1050 & 0.9284$\pm$0.0629 & 0.8953$\pm$0.0713\\
			& E2E-VarNet & 38.8871$\pm$3.5091  & 36.5089$\pm$2.9326 & 0.9359$\pm$0.0628  & 0.9131$\pm$0.0702\\
			& ReVarNet & 39.0496$\pm$3.6041 & 36.7979$\pm$3.0353 & 0.9366$\pm$0.0626 & 0.9145$\pm$0.0700\\
			& MeDL-Net & 38.6832$\pm$3.8629 & 36.4997$\pm$3.3799 &0.9332$\pm$0.0635 & 0.9161$\pm$0.0694\\
			& vSharp &39.2502$\pm$3.7787 & 36.9046$\pm$3.4141 & 0.9374$\pm$0.0617 & 0.9171$\pm$0.0699\\
			& MCU-Net (ours) & \textbf{39.4574$\pm$3.7197} & \textbf{37.4105$\pm$3.1319} & \textbf{0.9394$\pm$0.0616} & \textbf{0.9202$\pm$0.0686}\\
			\bottomrule
		\end{tabular}
	}
\end{table*}

\begin{table*}[!t]
	\caption{Computational costs of different DUNs trained on the coronal PD sequence.}\label{flops}
	\centering
	\resizebox{1.\columnwidth}{!}{
	\begin{tabular}{lllll} 
		\toprule
		Networks & Cascades & Average inference time (s) & FLOPs (G) & Memory (GB) \\
		\hline 
		D5C5 &7 &0.35& 186.55 & 2.70\\
		ISTA-Net &9 &0.35& 120.13 & 4.02\\
		VS-Net &10 &0.37& 265.89 & 3.70\\
		E2E-VarNet &12& 0.39& 187.39 &6.65\\
		ReVarNet &15& 0.48& 204.87 &24.23\\
		Medl-Net & 9 &0.40& 189.25 & 6.26 \\
		vSharp &12 & 0.50 & 609.63 & 15.41\\
		MCU-Net (ours) &10 &  0.46 & 213.07 & 9.85\\
		\bottomrule
	\end{tabular}}
\end{table*}

\begin{figure*}[!t]	
	\begin{minipage}[b]{1\linewidth}
		\centering
		\centerline{\includegraphics[width=12.8cm]{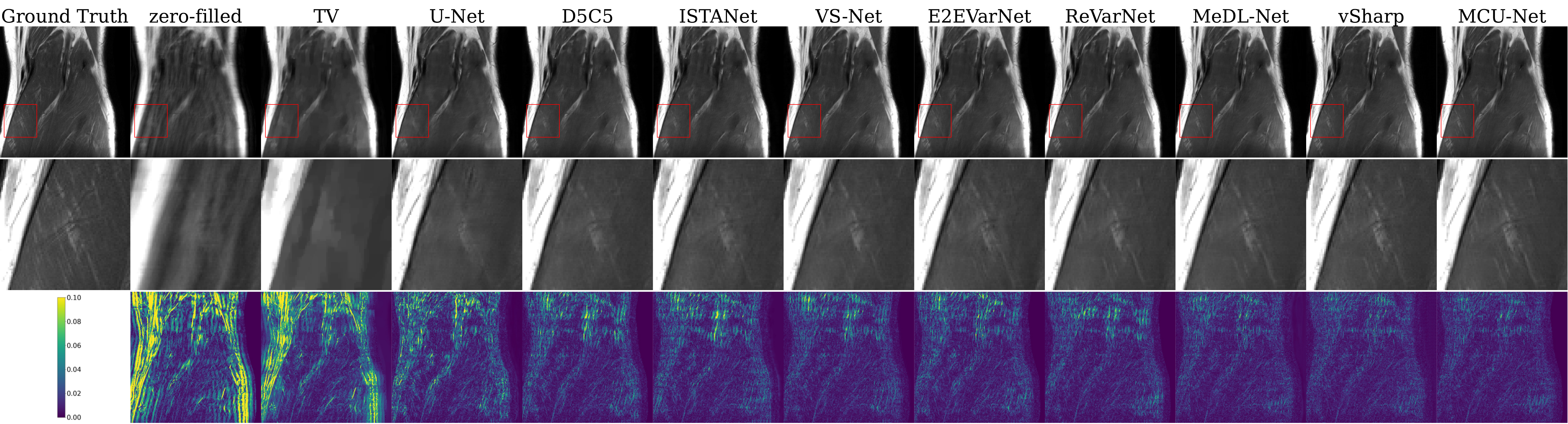}}
		\begin{center}
			\vspace{-0.1cm}
			\footnotesize coronal PD 4$\times$ acceleration
		\end{center}
	\end{minipage}
	\begin{minipage}[b]{1\linewidth}
		\centering
		\centerline{\includegraphics[width=12.8cm]{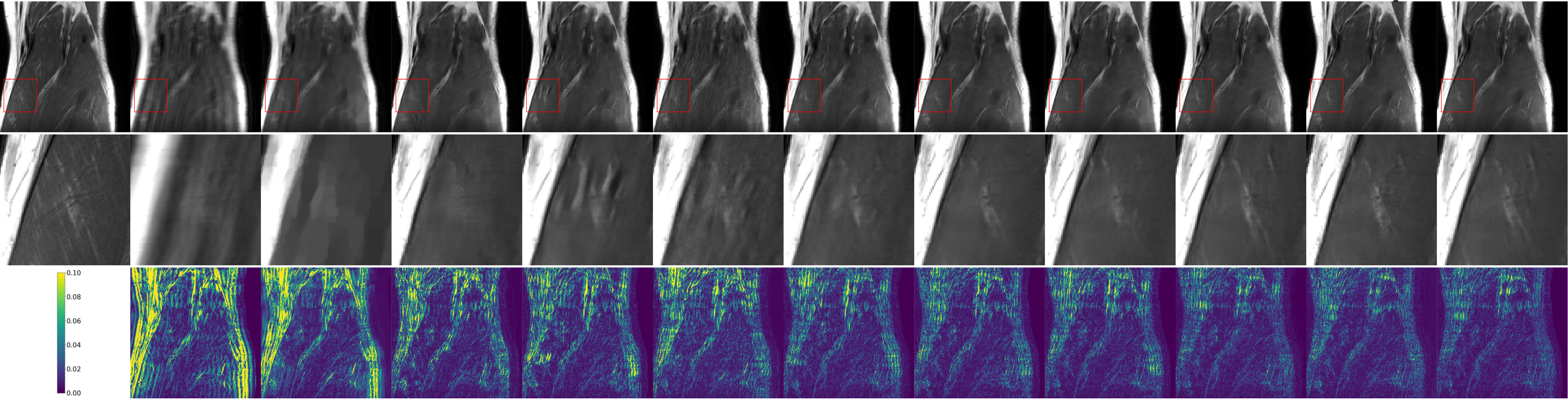}}
		\begin{center}
			\vspace{-0.1cm}
			\footnotesize coronal PD 6$\times$ acceleration
		\end{center}
	\end{minipage}
	\begin{minipage}[b]{1\linewidth}
		\centering
		\centerline{\includegraphics[width=12.8cm]{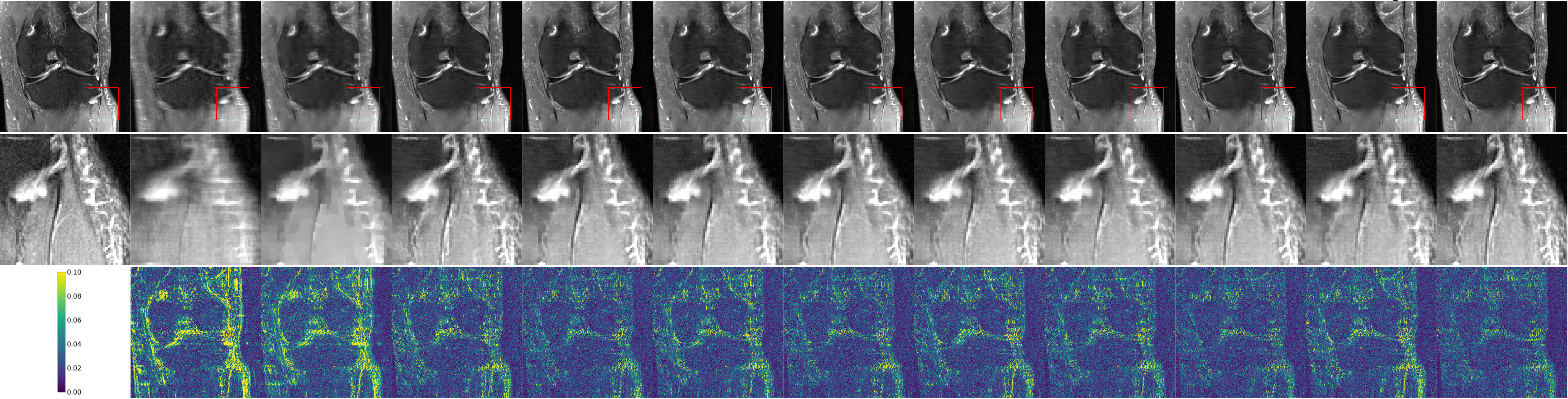}}
		\begin{center}
			\vspace{-0.1cm}
			\footnotesize coronal PDFS 4$\times$ acceleration
		\end{center}
	\end{minipage}
	\begin{minipage}[b]{1\linewidth}
		\centering
		\centerline{\includegraphics[width=12.8cm]{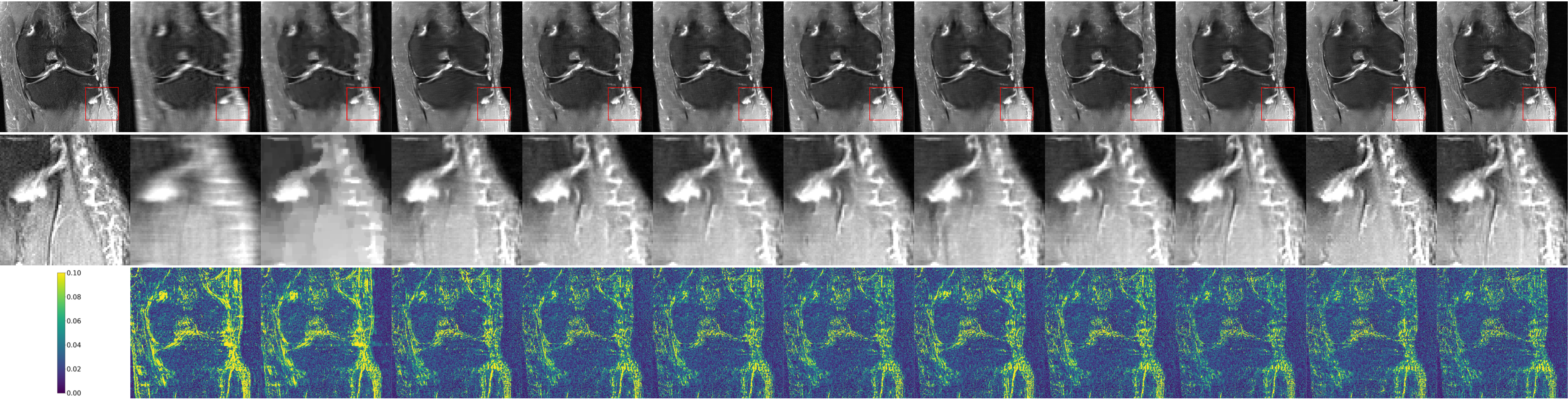}}
		\begin{center}
			\vspace{-0.1cm}
			\footnotesize coronal PDFS 6$\times$ acceleration
		\end{center}
	\end{minipage}
	\caption{Examples of reconstructed images of different methods. The first rows in each subfigures are the reconstructed images, the second rows are the zoomed details in the red boxes and the third rows are the corresponding error maps. Methods from left to right:1. ground truth; 2. zero-filled; 3. TV; 4. D5C5; 5. U-Net; 6. ISTA-Net; 7. VS-Net; 8. E2E-VarNet; 9. ReVarNet; 10. MeDL-Net; 11. vSharp; 12. MCU-Net (ours). 
		\label{fig:3}}
\end{figure*}

\begin{figure*}[!t]	
	\begin{minipage}[b]{1\linewidth}
		\centering
		\centerline{\includegraphics[width=13cm]{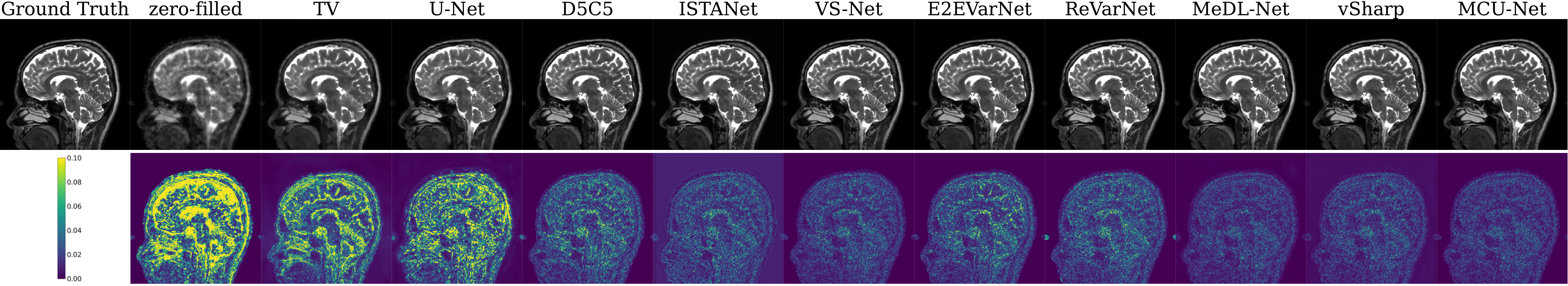}}
	\end{minipage}
	\caption{Examples of reconstructed brain MR images and the corresponding error maps. Methods from left to right: 1. ground truth; 2. zero-filled; 3. TV; 4. D5C5; 5. U-Net; 6. ISTA-Net; 7. VS-Net; 8. E2E-VarNet; 9. ReVarNet; 10. MeDL-Net; 11. vSharp; 12. MCU-Net (ours). 
		\label{fig:4}}
\end{figure*}

\begin{table*}[!t]
	\centering
	\caption{The quantitative results on the brain MRI dataset compared with the state-of-the-art methods.}\label{brainsota}
	\resizebox{1.\columnwidth}{!}{
	\begin{tabular}{llllll} 
		\toprule
		Method & average inference time (s) & FLOPs(G) & Memory (GB) & PSNR$\uparrow$ & SSIM$\uparrow$ \\
		\hline 
		zero-filled & - & - & - & 27.8175$\pm$0.1860 & 0.8437$\pm$0.0502 \\
		TV & - & - & - & 31.8378$\pm$1.8519& 0.8072$\pm$0.0731\\
		U-Net&0.13 & 12.37 & 0.86 & 30.8815$\pm$2.2455 & 0.8047$\pm$0.1872\\
		D5C5  &0.10 & 46.93 & 1.12& 36.9407$\pm$1.4127 & 0.9627$\pm$0.0125\\
		ISTA-Net&0.09& 30.17&1.63 & 36.9413$\pm$1.4091 & 0.8891$\pm$0.1192\\
		VS-Net &0.11&67.05&1.38& 39.6361$\pm$1.2324 & 0.9770$\pm$0.0106\\
		E2E-VarNet &0.14&47.20&2.35& 35.2120$\pm$4.6142 & 0.9428$\pm$0.0618\\
		ReVarNet &0.13&48.46&7.46& 37.5542$\pm$1.2790 & 0.9625$\pm$0.0169\\
		MeDL-Net &0.15 & 47.55 & 2.26 &39.5831$\pm$1.7288 & 0.9541$\pm$0.0614\\
		vSharp&0.17&153.19&4.92 & 38.4702$\pm$3.3468 & 0.8602$\pm$0.1659\\
		MCU-Net (ours) &0.19&53.61&3.60& \textbf{40.3497$\pm$1.11409} & \textbf{0.9789$\pm$0.0115}\\
		\bottomrule
	\end{tabular}}
\end{table*}

Experimental results conducted on the NYU dataset and the FastMRI dataset are presented in Table~\ref{sota}.
We tested $4\times$ and $6\times$ accelerations, indicating that we sampled one-quarter and one-sixth of the k-space data as input measurements, respectively. 
To evaluate the performance of various networks, we employed the most commonly used metrics: PSNR and SSIM.
In the table, the zero-filled method denotes the baseline results, where the undersampled raw data are filled with zeros.
TV denotes the conventional total variation algorithm \cite{ma2008efficient}.
U-Net denotes the basic U-Net structure \cite{zbontar2018fastmri}.
The number of pooling layers was set to 4, and the number of corresponding channels were set to 32, 64, 128, and 256.
D5C5\cite{schlemper2017deep}, ISTA-Net\cite{zhang2018ista}, VS-Net\cite{duan2019vs}, E2E-VarNet\cite{sriram2020end}, ReVarNet\cite{yiasemis2022recurrent}, MeDL-Net \cite{qiao2023medl} and vsharp \cite{yiasemis2023vsharp} are DUNs proposed to reconstruct MR images in recent years.
The computational costs of different networks are reported in Table~\ref{flops}.
For ReVarNet, it cannot be directly trained on our GPU (24GB).
Considering sufficient cascades is significant for achieving superior reconstruction performance, we preserved 15 casacades in the network and reduced the number of channels in ReVarNet from 64 to 32 to reduce its FLOPs to the same level as in our model.
Considering that D5C5 has lower FLOPs, we increased the number of cascades from five to seven in the experiments to improve its performance.
For the other networks, we tried our best to preserve the network parameters in their published papers.
As shown in Table \ref{flops}, ISTA-Net exhibits relatively lower FLOPs compared to other networks. 
However, we did not observe significant improvements in the results when increasing its parameters.
Hence, we retained the announced hyper-parameters, as in their study.
The results shown in the tables demonstrate that the proposed MCU-Net can achieve the best performance among the chosen methods with relatively low computational costs and limited number of COCAs.

Examples of the reconstructed images from different methods are plotted in Fig.~\ref{fig:3}. 
Here, we presented the results obtained from the coronal PD and PDFS sequences with 4-fold and 6-fold acceleration, while the results on the FastMRI dataset exhibited similar outcomes.
Fig.~\ref{fig:3} clearly demonstrates that our method excels in recovering image details with the fewest errors among all methods.
Besides, examples of reconstructed images as well as the corresponding error maps on the brain MRI data are shown in Fig.~\ref{fig:4}, and the quantitative results are presented in Table~\ref{brainsota}.
When applied to a smaller dataset with different anatomy and sampling trajectories, not all comparison methods achieve superior results. Conversely, our method consistently proves its superiority, emphasizing its effectiveness.

\subsection{Ablation studies}
\label{ablations}

\begin{table*}[!t]
	\caption{Results of ablation studies on coronal PD sequence with 4-fold and 6-fold acceleration rates.}\label{tab:ablation}
	\centering
	\resizebox{1.\columnwidth}{!}{
		\begin{tabular}{llllll} 
			\toprule
			\multirow{2}{*}{Method} & \multicolumn{2}{l}{PSNR$\uparrow$} & \multicolumn{2}{l}{SSIM$\uparrow$} & \multirow{2}{*}{FLOPs(G)}  \\\cline{2-5}
			& 4$\times$ & 6$\times$ & 4$\times$ & 6$\times$ &\\
			\hline 
			addition & 39.5436$\pm$3.2331 & 36.9604$\pm$3.0646 & 0.9547$\pm$0.0623 & 0.9347$\pm$0.0739 & 193.05 \\
			w/o GSAM & 40.5671$\pm$3.0143 & 37.8001$\pm$2.7495  & 0.9618$\pm$0.0614 & 0.9425$\pm$0.0691 & 201.07 \\
			w/o gate units & 40.9493$\pm$3.0194 & 38.0659$\pm$3.0012 & 0.9626$\pm$0.0604 & 0.9433$\pm$0.0721 & 205.14\\
			w/o RC & 41.0549$\pm$2.9581 & 38.2319$\pm$2.8691 & 0.9634$\pm$0.0610 & 0.9441$\pm$0.0678 & 212.73 \\
			w/o OC & 41.0099$\pm$3.4621 & 38.1916$\pm$2.9409 & 0.9629$\pm$0.0621  & 0.9438$\pm$0.0688  & 212.73\\
			w/o correction & 40.5920$\pm$3.1172 & 37.7134$\pm$3.3603 & 0.9582$\pm$0.0614 & 0.9396$\pm$0.0721 & 205.20\\
			w/o intermediate losses  & 40.9506$\pm$2.9205 & 38.0350$\pm$2.9829 & 0.9607$\pm$0.0605 & 0.9415$\pm$0.0699 & 213.07 \\
			original & \textbf{41.2003$\pm$2.9637} & \textbf{38.4734$\pm$2.8095} & \textbf{0.9641$\pm$0.0595} & \textbf{0.9455$\pm$0.0668} & 213.07  \\
			\bottomrule 
	\end{tabular}}
\end{table*}

\begin{table*}[!t]
	\caption{Reconstruction results with different inputs in the GSAMs.}\label{tab:ams}
	\centering
	\resizebox{1.\columnwidth}{!}{
		\begin{tabular}{lllllll} 
			\toprule
			\multirow{2}{*}{inputs} & \multirow{2}{*}{channel}  & \multicolumn{2}{l}{PSNR$\uparrow$} & \multicolumn{2}{l}{SSIM$\uparrow$} & \multirow{2}{*}{FLOPs(G)} \\\cline{3-6}
			&  &  4$\times$ & 6$\times$ & 4$\times$ & 6$\times$ &\\
			\hline 
			$x^k_c$ & 32 & 40.9493$\pm$3.0194 & 38.0659$\pm$3.0012 & 0.9626$\pm$0.0604 & 0.9433$\pm$0.0721 & 205.14\\
			$x^{k-1}_c, x^k_c$ & 32 & 40.9803$\pm$3.0296 &38.1735$\pm$3.0246 & 0.9629$\pm$0.0610 & 0.9439$\pm$0.0732 &207.58\\
			$x^0, x^{k-1}_c, x^k_c$ & 32 & 41.0221$\pm$3.0523 & 38.1785$\pm$2.9848 & 0.9634$\pm$0.0610& 0.9441$\pm$0.0727 & 208.93\\
			$x^0, x^{k-2}_c, x^{k-1}_c, x^k_c$ & 32 & 41.0672$\pm$3.0063 & 38.1687$\pm$3.0691 & 0.9633$\pm$0.0610 & 0.9440$\pm$0.0732 &211.10\\
			$x^0, x^{k-2}_c, x^{k-1}_c, x^k_c$ & 64 & 41.0258$\pm$2.9613 & 38.1636$\pm$3.1465 &0.9634$\pm$0.0610 & 0.9437$\pm$0.0759 &221.14\\
			$x^0, x^{k-3}_c, x^{k-2}_c, x^{k-1}_c, x^k_c$ & 32 & 40.9556$\pm$3.0352 & 38.2047$\pm$3.1269 & 0.9629$\pm$0.0606 &  0.9439$\pm$0.0747 &213.01\\
			$x^0, x^{k-3}_c, x^{k-2}_c, x^{k-1}_c, x^k_c$& 64 & 41.1108$\pm$3.0005 &38.2279$\pm$2.9977 & 0.9636$\pm$0.0600 &  0.9443$\pm$0.0723 &224.94\\
			gate-controlled features & 16 &\textbf{41.2003$\pm$2.9637} & \textbf{38.4734$\pm$2.8095} & \textbf{0.9641$\pm$0.0595} & \textbf{0.9455$\pm$0.0668} & 213.07  \\
			\bottomrule 
	\end{tabular}}
\end{table*}
We designed ablation studies to demonstrate the improvements achieved by the proposed modules.
The experimental results obtained from the coronal PD sequence are presented in Table~\ref{tab:ablation}.
Here, we evaluated the following networks:
\begin{itemize}
	\item addition: GSAMs and CMs are removed from MCU-Net. The mean values of $x_s^k$ and $x_l^k$ are directly fed to the next COCAs.
	\item w/o GSAM: Only the GSAMs are removed from MCU-Net. The mean values, instead of the weighted images are fed to the CMs, and the corrected images in CMs directly replaced the fused images (i.e., without the OC-based weighting process).
	\item w/o gate units: $\mathcal{M}^k_R$ and $\mathcal{M}^k_O$ are learned from $x_c^k$ using convolutions, while all the gate units are removed. 
	\item w/o RC: The GSAMs only learn the OC maps, and mean values of IRs are fed to the CMs.
	\item w/o OC: The GSAMs only learn the RC maps, and the corrected images in CMs directly replaced the weighted images.
	\item w/o correction: only the CMs are removed from MCU-Net, which means that the weighted IRs from GSAMs are fed to the next COCAs.
	\item w/o intermediate losses: only the final results are fed to the loss function (without the IRs).
	\item original: The proposed MCU-Net.
\end{itemize}

The 'addition' method represents the simplest parallel network, where the mean values of the intermediate results from $N_s$ and $N_l$ are directly fed to the $(k+1)$th COCA.
Compared with MCU-Net, the improvements brought by all the proposed modules are significant.
When the GSAMs are removed from MCU-Net, the subnetworks contribute equally to the IRs. 
However, because different subnetworks have varying abilities to reconstruct images, this process cannot guarantee improvements in fused results. 
Furthermore, directly replacing fused IRs with corrected images in CMs without learned OC maps could potentially introduce new errors.
As a result, removing GSAMs from MCU-Net exhibits inferior performance.
When the gate units in GSAMs are removed, the confidences maps are learned solely from the IRs in the same COCA. Compared with MCU-Net, we can observe the benefits of adaptively introducing auxiliary information from various COCAs.

Then, we evaluated the individual contribution of RC and OC maps.
As illustrated in the table, varying degrees of deterioration were observed in 'w/o RC' and 'w/o OC' methods, underscoring the necessity of two-stage weighting process.
Similarly, removing the CMs from MCU-Net limited the network flexibility, impairing its ability to effectively handle low-OC pixels, which led to suboptimal results. 
The intermediate losses imposed on each COCA helps to constrain the functionality of the confidence maps, as inaccurate confidence maps result in increased losses.
When these intermediate losses are removed, only the final output is adopted to guide network training. This results in slower convergence and a deterioration in the performance of reconstructed images.

Moreover, we observed that the designed modules are easy to implement, resulting in significant improvements, while the increase in FLOPs is negligible. 
Compared to the naive 'addition' methods, the MCU-Net exhibited an approximate 10\% increase in FLOPs but achieved substantial improvements in results, demonstrating its effectiveness.

\begin{table*}[!t]
	\caption{Comparisons with non-parallel networks on coronal PD sequence with 4-fold and 6-fold acceleration rates.}\label{tab:ablation2}
	\centering
	\resizebox{1.\columnwidth}{!}{
		\begin{tabular}{lllllll} 
			\toprule
			\multirow{2}{*}{Method} & \multirow{2}{*}{Modifications to hyperparameters} & PSNR$\uparrow$ & & SSIM$\uparrow$ & & \multirow{2}{*}{FLOPs(G)} \\\cline{3-6}
			& & 4$\times$ & 6$\times$ & 4$\times$ & 6$\times$ &\\
			\hline 
			L-Net & cascades: $10\rightarrow15$ & 40.2784$\pm$2.9231 & 37.1909$\pm$2.8541 & 0.9558$\pm$0.0591 & 0.9296$\pm$0.0684 & 198.18 \\
			L-Net & channels: $16\rightarrow20$ & 40.0213$\pm$2.9835 &37.3131$\pm$2.8203  & 0.9556$\pm$0.0600 & 0.9296$\pm$0.0691 & 193.08 \\
			S-Net & cascades: $10\rightarrow15$ channels: $32\rightarrow48$ & 38.7506$\pm$2.9279  & 35.9477$\pm$2.8162 & 0.9485$\pm$0.0590 & 0.9234$\pm$0.0679 & 228.93 \\
			L+S-Net & - & 40.3504$\pm$2.9365 & 37.6080$\pm$2.8360 & 0.9587$\pm$0.0594 & 0.9366$\pm$0.0712 & 201.07 \\
			MCU-Net-2 & number of COCAs: $10\rightarrow2$ &39.7208$\pm$2.9184&36.9038$\pm$2.8647&0.9571$\pm$0.0601&0.9349$\pm$0.0658&42.21\\
			MCU-Net-4& number of COCAs: $10\rightarrow4$ &40.5196$\pm$2.9085&37.6705$\pm$2.7618&0.9609$\pm$0.0568&0.9404$\pm$0.0649&84.92\\
			MCU-Net-6& number of COCAs: $10\rightarrow6$ &40.8702$\pm$2.9547&38.0786$\pm$2.8940&0.9616$\pm$0.0581&0.9431$\pm$0.0650&127.64\\
			MCU-Net-8& number of COCAs: $10\rightarrow8$ &41.0982$\pm$2.9103&38.3074$\pm$2.8510&0.9629$\pm$0.0591&0.9442$\pm$0.0648&170.35\\
			MCU-Net (ours) & - &\textbf{41.2003$\pm$2.9637} & \textbf{38.4734$\pm$2.8095} & \textbf{0.9641$\pm$0.0595} & \textbf{0.9455$\pm$0.0668} & 213.07  \\
			
			\bottomrule 
		\end{tabular}
	}
\end{table*}

\begin{figure}[!t]	
	\begin{minipage}[b]{1\linewidth}
		\centering
		\centerline{\includegraphics[width=14cm]{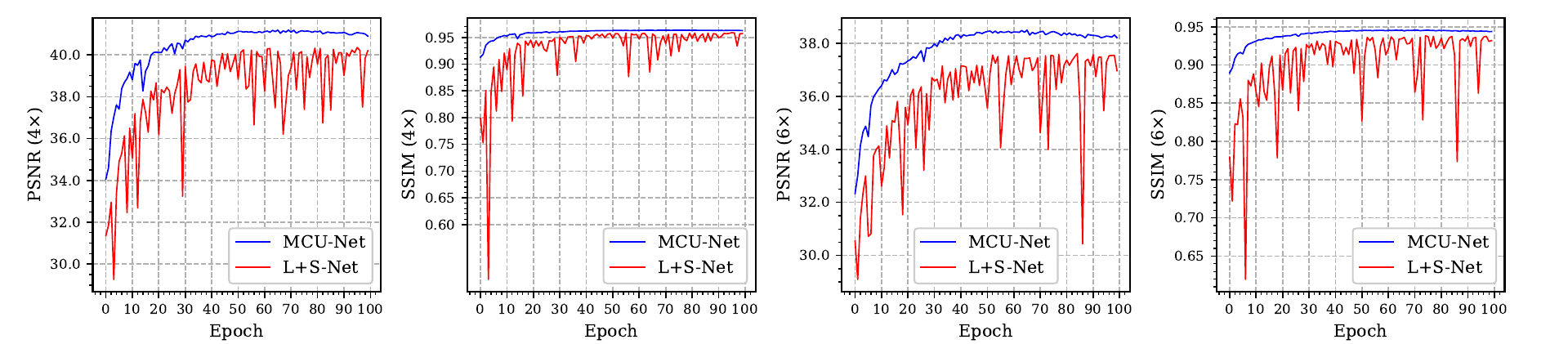}}
	\end{minipage}
	\caption{The test PSNR and SSIM results of L+S-Net and the proposed MCU-Net at 4-fold and 6-fold acceleration.
		\label{fig:5}}
\end{figure}

\begin{figure}[!t]	
	\begin{minipage}[b]{1\linewidth}
		\centering
		\centerline{\includegraphics[width=14cm]{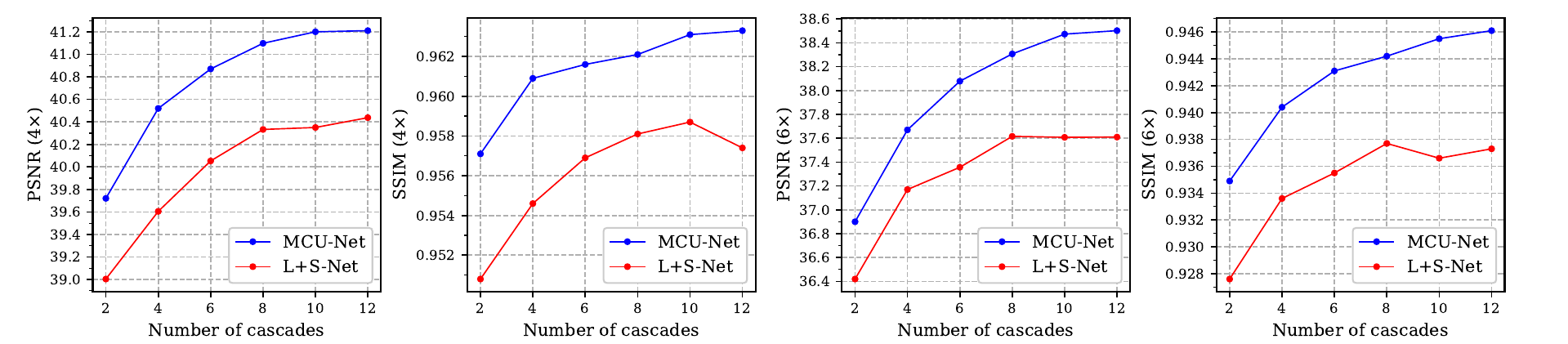}}
	\end{minipage}
	\caption{The test PSNR and SSIM results of L+S-Net and the proposed MCU-Net at 4-fold and 6-fold acceleration with different number of cascades.
		\label{fig:5_1}}
\end{figure}

To further emphasize the improvements introduced by GSAMs, we designed ablation networks in which the attention module directly learns from enriched inputs comprising intermediate results across different COCAs. 
As illustrated in Table \ref{tab:ams}, rather than being adaptively modulated by gate units, the enriched inputs are manually selected, concatenated, and fed into the convolutional layers in attention modules to incorporate both short- and long-term information. 
Additionally, we also experimented with increasing the number of channels in the convolutions. 
We observed that both enriched inputs and an increased number of channels generally lead to improved performance, indicating that incorporating short- and long-term information from various COCAs can enhance the robustness of the learned confidence maps. However, despite the higher FLOPs, these networks still exhibited inferior performance compared to MCU-Net. This experiment could further demonstrate the effectiveness of the GSAMs.

\subsection{Comparisons with naive cascades}

All the ablation studies utilize the proposed parallel network structure which incorporate prior knowledge in different ways.
To further demonstrate the efficacy of the proposed method compared to a non-parallel structure that employs naive cascades, we conducted comparisons with two types of networks: (a) the removal of one OSN from MCU-Net, and (b) the alternative optimization of the multi-prior objective function.

For type (a), the FLOPs of these networks are reduced due to the removal of components in the OSN and the modules used to fuse the OSNs. 
To mitigate the interference caused by varying FLOPs, we augmented the parameters of type (a) networks to ensure they have similar FLOPs with the proposed network. 
The designed networks as well as the corresponding modifications to the hyperparameters are listed in Table~\ref{tab:ablation2}.
For example, the first L-Net denotes a DUN constructed with the naive cascades from $N_l$, and the number of cascades was increased to 15 from 10.
For the S-Net, which is constructed with the naive cascades in $N_s$, because its FLOPs is relatively lower, merely increasing the cascade or depth makes the network difficult to train.
Therefore, we simultaneously augmented the number of channels in the learned transform from 32 to 48 and the number of cascades from 10 to 15.

For type (b), we designed L+S-Net, which solves the following problem:
\begin{equation}\label{linearobj}
	\hat{x}=\mathop{\arg\min}\limits_{x}\|\mathcal A x-y\|_{2}^{2}+\lambda_1 \|\Phi(x)\|_1+\lambda_2 \|\tau(x)\|_*
\end{equation}
and the iterative solution steps are given by:
\begin{equation}
	\begin{aligned}
		\left\{ 
		\begin{array}{l}
			r_l^k = \lambda_2^k N_{lu}(x^{k-1})\\
			x_l^k=x^{k-1}-\beta^k (\mathcal A^{*}(\mathcal A(x^{k-1})-y)+2r_l^k)\\
			x^k=\mathcal{\widetilde{G}}(soft(\mathcal G(x_l^k), \lambda_1^k))
		\end{array} \right.
	\end{aligned}
\end{equation}
Essentially, the output of the low-rank-inspired branch is fed to the $N_s$ branch in L+S-Net, and different priors contribute to the results in an alternating form. 
Conversely, the outputs of $N_l$ and $N_s$ are directly fed into the GSAMs in our method.
As shown in the table, our method outperforms the other methods with $4\times$ and $6\times$ accelerations.
Besides, we plotted the PSNR and SSIM test results of L+S-Net and our method across epochs, as shown in Fig~\ref{fig:5}.
We observed that by utilizing COCAs, the convergence of the proposed MCU-Net is faster and more stable. 

Moreover, we evaluated the network performance of MCU-Net with reduced number of COCAs, and the results are shown in Table~\ref{tab:ablation2} (The last five methods).
We observed that with reduced number of COCAs, the MCU-Net-4 has already achieved better results than the naive DUN.
Fig.~\ref{fig:5_1} exhibits the comparison between L+S-Net and MCU-Net with different number of cascades, and the improvements brought by introducing COCA into DUNs are significant. 
In conclusion, these experimental results demonstrated that with the proposed collaborative strategy, the reconstruction performance are significantly improved with faster convergence and lower FLOPs compared with the naive DUN without adopting COCAs.

\subsection{Applying COCA to different DUNs for MRI Reconstruction}

\begin{table*}[!t]
	\caption{Results of applying our method to other DUNs on coronal PD sequence with 4-fold and 6-fold acceleration rates.}\label{tab:apply}
	\centering
	\resizebox{1.\columnwidth}{!}{
		\begin{tabular}{lllllll} 
			\toprule
			\multirow{2}{*}{Method} & \multirow{2}{*}{Modifications to hyperparameters} & \multicolumn{2}{l}{PSNR$\uparrow$} & \multicolumn{2}{l}{SSIM$\uparrow$} &\multirow{2}{*}{FLOPs(G)}\\\cline{3-6}
			& &  4$\times$ & 6$\times$ & 4$\times$ & 6$\times$&\\
			\hline 
			VS-Net & - & 40.1649$\pm$2.8767 & 37.0604$\pm$2.7927 & 0.9591$\pm$0.0569 & 0.9358$\pm$0.0648 & 265.89\\
			VS\&L & channels: $64\rightarrow32$ & 41.1315$\pm$2.9537 & 38.3883$\pm$2.7927 & 0.9634$\pm$0.0621 & 0.9452$\pm$0.0687 & 212.07\\
			VS-Net+ & cascades: $10\rightarrow15$ & 40.4903$\pm$2.8800 & 37.4789$\pm$2.7207 & 0.9608$\pm$0.0558 & 0.9388$\pm$0.0651 & 398.83\\
			VS\&L+ & - & 41.1743$\pm$2.9580 & 38.3136$\pm$3.0333 & 0.9638$\pm$0.0607 & 0.9451$\pm$0.0679 & 410.13\\
			
			\hline 
			E2E-VarNet &- & 39.3592$\pm$2.9368 & 36.8949$\pm$3.3872 & 0.9542$\pm$0.0662 & 0.9342$\pm$0.0775 & 187.87 \\
			E2E-VarNet\&L& channels: $16\rightarrow8$& 41.0779$\pm$2.9625 & 38.3053$\pm$2.8231 & 0.9633$\pm$0.0593 & 0.9448$\pm$0.0662 & 210.81\\
			E2E-VarNet+ &cascades: $12\rightarrow14$, channels: $18\rightarrow22$& 38.9097$\pm$3.4703 & 36.0596$\pm$2.9721 & 0.9511$\pm$0.0785 & 0.9273$\pm$0.0740 & 326.79\\
			E2E-VarNet\&L+ & - & 41.0897$\pm$2.9352 & 38.3311$\pm$2.8092 & 0.9633$\pm$0.0585 & 0.9444$\pm$0.0667 & 321.85\\
			
			\bottomrule  
		\end{tabular}
	}
\end{table*}

\begin{figure*}[t]	
	\begin{minipage}[b]{1\linewidth}
		\centering
		\centerline{\includegraphics[width=14cm]{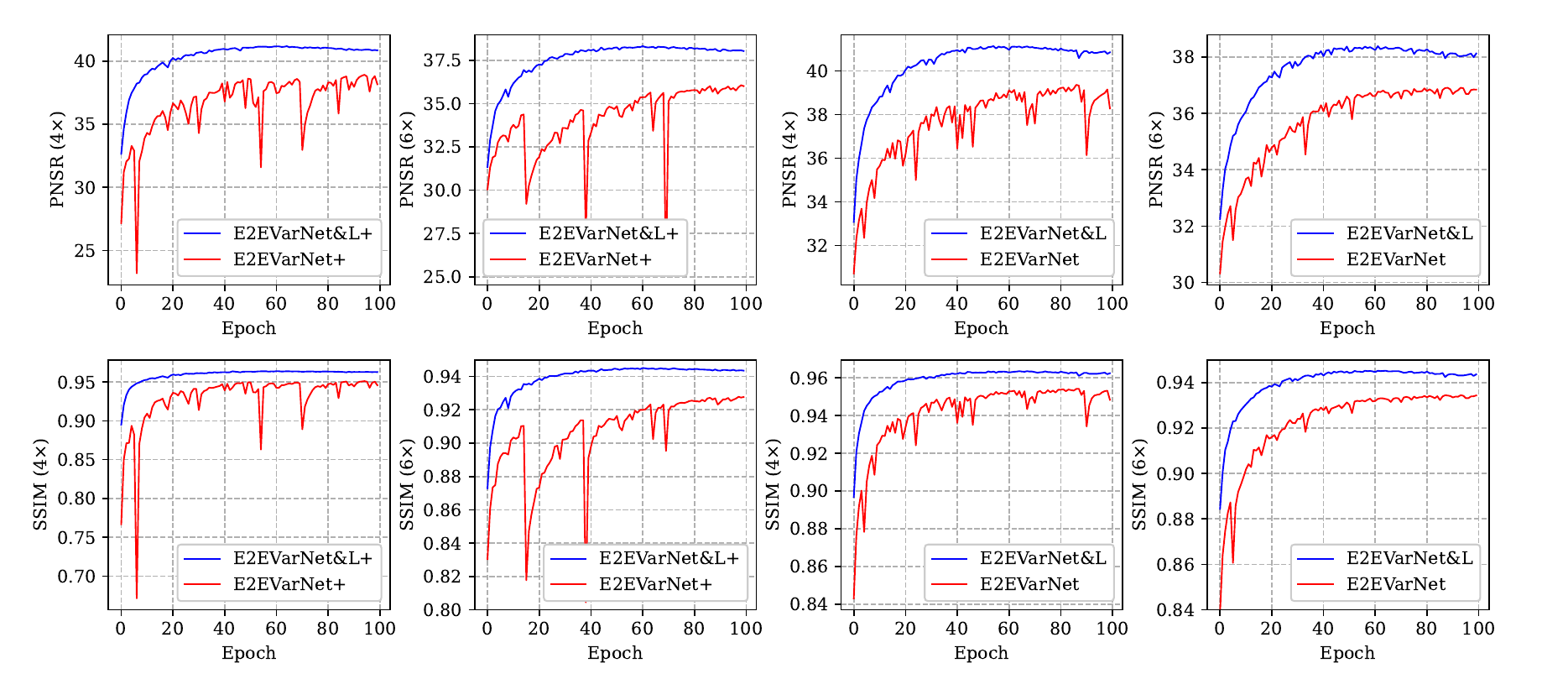}}
	\end{minipage}
	\caption{PSNR and SSIM results of E2E-VarNet-based networks at 4-fold and 6-fold acceleration. 
		\label{fig:6}}
\end{figure*}

Another advantage of the proposed collaborative network is that it imposes very few restrictions on the structure of the OSNs.
In other words, it can be easily applied to different optimization-inspired network structures to enhance performance.
In this section, we first conducted experiments to enhance VS-Net and E2E-VarNet with COCAs.
Considering that directly applying COCA to other DUNs will inevitably increase the computational costs, we tried two ways to unify the FLOPs of different networks: enlarging the original networks or shrinking the collaboration-based networks.
The experimental results on the coronal PD sequence are presented in Table~\ref{tab:apply}.
To apply COCAs to VS-Net, we replaced $N_s$ in MCU-Net with the cascades in VS-Net, labeled as 'VS\&L+'.
Meanwhile, we increased the cascade number in VS-Net from 10 to 15, labeled as 'VS-Net+', to compare with 'VS\&L+'.
As shown in the table, improvements can be observed at $4\times$ and $6\times$ accelerations in 'VS\&L+'.
In another group of comparison, we shrank the designed network, where the number of channels in the convolutional layers was reduced to 32 from 64, to compare with the original VS-Net.
The method labeled as 'VS\&L' also outperforms the original VS-Net.
Similarly, we observed improvements when applying our methods to E2E-VarNet.
In addition, we plotted the PSNR and SSIM results of E2E-VarNet-based methods at different epochs in Fig.~\ref{fig:6}.
The figure clearly showed that the applying COCA in DUNs (the blue lines) accelerates the network convergence and improves the performance.

\begin{table*}[!t]
	\caption{Experimental Results of adopting same structures in the OSNs.}\label{tab:apply2}
	\centering
	\resizebox{1.\columnwidth}{!}{
		\begin{tabular}{lllllll} 
			\toprule
			\multirow{2}{*}{structure} & \multirow{2}{*}{Method} & \multicolumn{2}{l}{PSNR$\uparrow$} & \multicolumn{2}{l}{SSIM$\uparrow$} & \multirow{2}{*}{FLOPs(G)}\\\cline{3-6}
			& & 4$\times$ & 6$\times$ & 4$\times$ & 6$\times$&\\
			\hline 
			\multirow{3}{*}{original} & ISTA-Net & 39.2910$\pm$2.8423 & 35.5220$\pm$3.0915 & 0.9471$\pm$0.0567 & 0.9138$\pm$0.0717 & 120.13\\
			& VS-Net & 40.1649$\pm$2.8767 & 37.0604$\pm$2.7927 & 0.9591$\pm$0.0569 & 0.9358$\pm$0.0648 & 265.89\\
			& E2E-VarNet & 39.3592$\pm$2.9368 & 36.8949$\pm$3.3872 & 0.9542$\pm$0.0662 & 0.9342$\pm$0.0775 & 187.87\\
			\hline 
			\multirow{2}{*}{enlarged} & VS-Net & 40.7450$\pm$2.9355 & 37.8016$\pm$2.8010 & 0.9620$\pm$0.0586&0.9411$\pm$0.0655& 531.79\\
			& E2EVarNet &38.9569$\pm$3.8205&35.6815$\pm$4.0284&0.9513$\pm$0.0741 & 0.9181$\pm$0.0836 &313.12\\
			\hline
			\multirow{3}{*}{collaborative}& VS\&VS & 41.0567$\pm$3.0592 & 38.1201$\pm$3.0491 & 0.9628$\pm$0.0658 & 0.9440$\pm$0.0669 & 300.44\\
			& ISTA\&ISTA & 40.8609$\pm$2.9376 & 37.9620$\pm$2.8670 & 0.9620$\pm$0.0578 & 0.9424$\pm$0.0662 &158.39\\
			& Var\&Var &41.1595$\pm$2.9590& 38.2810$\pm$2.9398 & 0.9635$\pm$0.0621 & 0.9448$\pm$0.0704 &333.13\\
			\bottomrule  
		\end{tabular}
	}
\end{table*}

\begin{figure*}[t]	
	\begin{minipage}[b]{1\linewidth}
		\centering
		\centerline{\includegraphics[width=13cm]{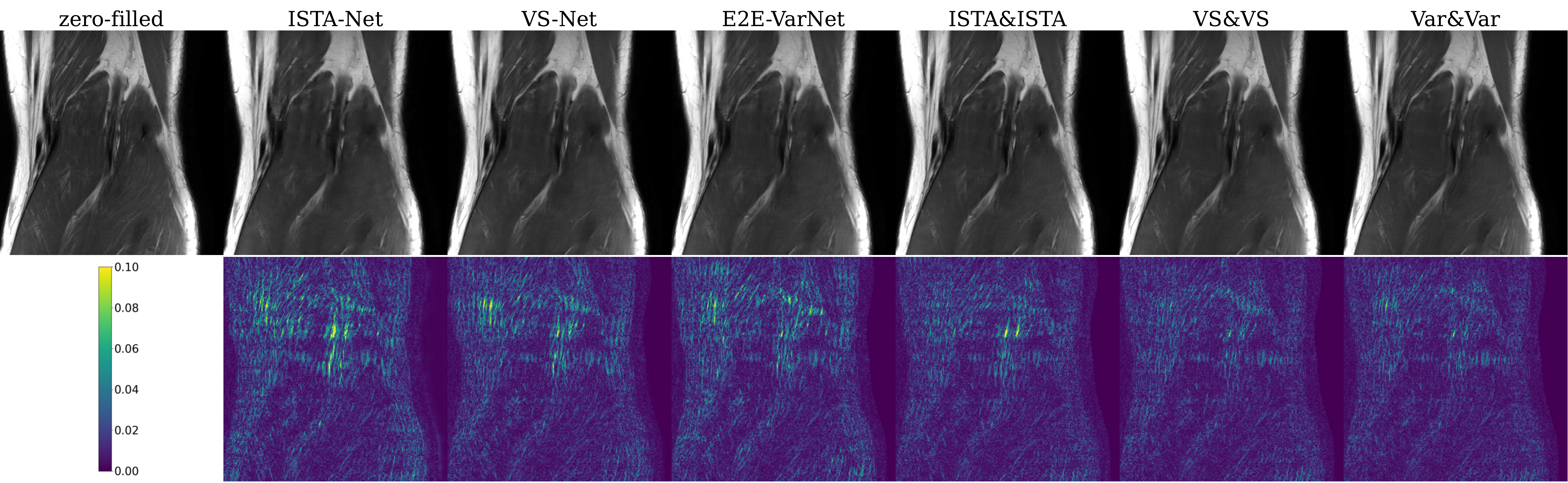}}
	\end{minipage}
	\caption{Examples of reconstructed MR images and the corresponding error maps from collaborative networks and the original counterparts at 4-fold acceleration. 
		\label{fig:7}}
\end{figure*}

Table~\ref{tab:apply2} presents the results when both OSNs in MCU-Net are replaced with the same structure, and significant improvements are similarly observed in collaborative networks compared to the original ones.
When the OSNs share the same structure, adopting COCAs can be regarded as a superior strategy for enlarging DUNs, rather than simply stacking more cascades. 
We also experimented with adding more cascades to the original networks shown in Table~\ref{tab:apply2} (i.e., the enlarged networks).
While VS-Net demonstrates better results than the original version, the reconstruction performance of E2EVarNet deteriorates, and ISTA-Net fails to converge. 
Conversely, applying COCA stabilizes network training and consistently yields better results. 
These findings further highlight the versatility of our network design strategy, making it easier to apply to different methods without the constraint of strictly unrolling cascades from various prior-based algorithms. 
However, these methods either have high FLOPs or yield inferior results compared to MCU-Net, showcasing the advantage of leveraging complementary information from different optimization-inspired processes. 
Additionally, the reconstructed image examples from the original and collaborative networks, illustrated in Fig.~\ref{fig:7}, further demonstrate the effectiveness of collaborative networks, which exhibit superior suppression of reconstruction errors.

\section{Discussion}
In the context of MRI reconstruction and related fields \cite{mokry2023improving, lin2018efficient, dong2020low}, incorporating various priors into the objective function is a common practice, as it effectively narrows the solution space. 
For example, in \cite{chen2024thick}, the authors exploited the low-rankness and sparsity prior from image and cloud components, respectively, in a cloud removal task, which was solved with HQS algorithm.
The authors in \cite{chen2023combining} combined low-rankness and deep plug-and-play
priors for snapshot compressive imaging.
However, most methods alternatively minimize the multi-prior based optimization problem \cite{ke2021learned, huang2021deep}, and to the best of our knowledge, few of the related methods have further explored the optimal method of unfolding multi-prior based optimizations into deep networks.
The authors in \cite{xu2024attention} employed HQS algorithm to iteratively solve multi-prior problem.
They utilized a weighted addition operator to share low-rank and sparse information, which allows for further optimization.
Conversely, we designed different optimization-inspired subnetworks, forming a parallel network structure.
The primary distinction in our method is the introduction of a two-stage input-level weighting procedure in each COCA, which aims to fully exploit the underlying complementarity and enhance network convergence. 
We introduced two confidence maps to evaluate the quality of IRs in each COCA.
The proposed method is particularly crucial in highly ill-posed problems, as inaccuracies in shallow cascades could be amplified and impact subsequent training in the absence of GSAMs and DMs.
Our experimental results have demonstrated the efficiency of the proposed COCAs when applying to MRI reconstruction, showcasing significant improvements over naive cascades. Moreover, the proposed method offers enhanced versatility, allowing for improved reconstruction performance even when only a single prior is available.

In this study, we designed a spatial attention-based module for precise integration of intermediate results from different subnetworks.
The attention mechanism has been widely applied in related reconstruction networks \cite{yuan2020sara, wu2019self,zhou2021spatial}.
For example, HUANG et al. \cite{huang2019mri} introduced a variant of U-Net with a channel-wise attention module to suppress irrelevant features.
In \cite{liu2022coil}, the purpose of attention modules was to adaptively capture local contextual cues.
The authors in \cite{liu2022dual} applied the squeeze-and-excitation (SE) attention to adaptively weight the channels.
In \cite{zhou2023rnlfnet}, the authors captured long-range spatial dependencies in the frequency domain by combining self-attention with Fourier transform.
In comparison, we proposed a unique two-stage optimization strategy based on attention in an element-wise manner, and the experimental results demonstrated the learned RC and OC maps facilitate each COCA in approaching the target images more rapidly, thereby optimizing overall network convergence.

Recurrent Neural networks \cite{das2023recurrent} and LSTM-based networks have also been introduced into the task of MRI reconstruction. 
For example, ReVarNet \cite{yiasemis2022recurrent} was constructed with recurrent blocks, which performs iterative optimization in k-space.
The authors in \cite{chen2022pyramid} designed pyramid convolutional RNN, reconstructing MR images in a coarse-to-fine manner.
In \cite{song2021memory, li2024gates}, LSTM was adopted to construct DUN, while the proximal gradient descent was further aggregated into ConvLSTM blocks in \cite{li2024gates}.
Conversely, we designed gate units to extract key information during the training process, which is subsequently used to evaluate the confidence maps of IRs. Compared to existing methods, our approach provides a novel idea for enhancing information interactions among cascades while offering greater flexibility in network design.

In addition, a limitation of this study is that the network achieves effective collaboration primarily in the image domain, while the potential for cross-domain complementarity remains underutilized. 
In future studies, our method can be extended to explore complementary information from the image and frequency domain, further demonstrating its effectiveness.

\section{Conclusion}
In this study, we proposed a collaborative DUN constructed with COCAs for MRI reconstruction.
We designed RC and OC map-based attention and correction modules to maximize the contributions of different OSNs in their respective areas of expertise.
Compared to naive cascades, the proposed method demonstrates significant improvements even with a greatly reduced number of COCAs. 
Experimental results on various datasets highlight the superiority of our approach. When applied to different optimization-inspired structure, significant improvements were observed without additional increments in FLOPs.

\section{Acknowledgements}
This work was supported by the National Natural Science Foundation of China [Nos. 62331008,  62027827, 62221005 and 62276040], National Science Foundation of Chongqing [Nos. 2023NSCQ-LZX0045 and CSTB2022NSCQ-MSX0436], Chongqing University of Posts and Telecommunications ph.D. Innovative Talents Project [BYJS202211].

\bibliographystyle{elsarticle-num}
\bibliography{sn-bibliography}

\begin{thebibliography}{10}
\expandafter\ifx\csname url\endcsname\relax
  \def\url#1{\texttt{#1}}\fi
\expandafter\ifx\csname urlprefix\endcsname\relax\def\urlprefix{URL }\fi
\expandafter\ifx\csname href\endcsname\relax
  \def\href#1#2{#2} \def\path#1{#1}\fi

\bibitem{tu2024novel}
Y.~Tu, S.~Lin, J.~Qiao, K.~Hao, Y.~Zhuang, A novel dual-branch alzheimer’s
  disease diagnostic model based on distinguishing atrophic patch localization,
  Applied Intelligence (2024) 1--21.

\bibitem{chen2024general}
Y.~Chen, X.~Yang, X.~Yue, X.~Lin, Q.~Zhang, H.~Fujita, A general
  variation-driven network for medical image synthesis, Applied Intelligence
  54~(4) (2024) 3295--3307.

\bibitem{yang2023deep}
H.~Yang, Z.~Wang, X.~Liu, C.~Li, J.~Xin, Z.~Wang, Deep learning in medical
  image super resolution: a review, Applied Intelligence 53~(18) (2023)
  20891--20916.

\bibitem{harisinghani2019advances}
M.~G. Harisinghani, A.~O’Shea, R.~Weissleder, Advances in clinical mri
  technology, Science Translational Medicine 11~(523) (2019) eaba2591.

\bibitem{larkman2007parallel}
D.~Larkman, R.~Nunes, Parallel magnetic resonance imaging, Phys Med Biol 52~(7)
  (2007) R15.

\bibitem{donoho2006compressed}
D.~L. Donoho, Compressed sensing, IEEE Transactions on information theory
  52~(4) (2006) 1289--1306.

\bibitem{pruessmann1999sense}
K.~Pruessmann, M.~Weiger, M.~Scheidegger, P.~Boesiger, Sense: sensitivity
  encoding for fast mri, Magn Reson Med 42~(5) (1999) 952--962.

\bibitem{griswold2002generalized}
M.~A. Griswold, P.~M. Jakob, R.~M. Heidemann, M.~Nittka, V.~Jellus, J.~Wang,
  B.~Kiefer, A.~Haase, Generalized autocalibrating partially parallel
  acquisitions (grappa), Magnetic Resonance in Medicine: An Official Journal of
  the International Society for Magnetic Resonance in Medicine 47~(6) (2002)
  1202--1210.

\bibitem{lustig2007sparse}
M.~Lustig, D.~Donoho, J.~M. Pauly, Sparse mri: The application of compressed
  sensing for rapid mr imaging, Magnetic Resonance in Medicine: An Official
  Journal of the International Society for Magnetic Resonance in Medicine
  58~(6) (2007) 1182--1195.

\bibitem{lustig2008compressed}
M.~Lustig, D.~L. Donoho, J.~M. Santos, J.~M. Pauly, Compressed sensing mri,
  IEEE signal processing magazine 25~(2) (2008) 72--82.

\bibitem{liu2018cs}
S.~Liu, J.~Cao, G.~Wu, H.~Liu, X.~Tan, X.~Zhou, Cs-mri reconstruction via
  group-based eigenvalue decomposition and estimation, Neurocomputing 283
  (2018) 166--180.

\bibitem{zhao2010low}
B.~Zhao, J.~P. Haldar, C.~Brinegar, Z.-P. Liang, Low rank matrix recovery for
  real-time cardiac mri, in: 2010 ieee international symposium on biomedical
  imaging: From nano to macro, IEEE, 2010, pp. 996--999.

\bibitem{hollingsworth2015reducing}
K.~Hollingsworth, Reducing acquisition time in clinical mri by data
  undersampling and compressed sensing reconstruction, Phys Med Biol 60~(21)
  (2015) R297.

\bibitem{qu2010iterative}
X.~Qu, W.~Zhang, D.~Guo, C.~Cai, S.~Cai, Z.~Chen, Iterative thresholding
  compressed sensing mri based on contourlet transform, Inverse Probl Sci En
  18~(6) (2010) 737--758.

\bibitem{voulodimos2018deep}
A.~Voulodimos, N.~Doulamis, A.~Doulamis, E.~Protopapadakis, Deep learning for
  computer vision: A brief review, Computational intelligence and neuroscience
  2018~(1) (2018) 7068349.

\bibitem{wang2020deepcomplexmri}
S.~Wang, H.~Cheng, L.~Ying, T.~Xiao, Z.~Ke, H.~Zheng, D.~Liang, Deepcomplexmri:
  Exploiting deep residual network for fast parallel mr imaging with complex
  convolution, Magnetic Resonance Imaging 68 (2020) 136--147.

\bibitem{akccakaya2019scan}
M.~Ak{\c{c}}akaya, S.~Moeller, S.~Weing{\"a}rtner, K.~U{\u{g}}urbil,
  Scan-specific robust artificial-neural-networks for k-space interpolation
  (raki) reconstruction: Database-free deep learning for fast imaging, Magnetic
  resonance in medicine 81~(1) (2019) 439--453.

\bibitem{mardani2018deep}
M.~Mardani, E.~Gong, J.~Y. Cheng, S.~S. Vasanawala, G.~Zaharchuk, L.~Xing,
  J.~M. Pauly, Deep generative adversarial neural networks for compressive
  sensing mri, IEEE transactions on medical imaging 38~(1) (2018) 167--179.

\bibitem{quan2018compressed}
T.~M. Quan, T.~Nguyen-Duc, W.-K. Jeong, Compressed sensing mri reconstruction
  using a generative adversarial network with a cyclic loss, IEEE transactions
  on medical imaging 37~(6) (2018) 1488--1497.

\bibitem{noor2024dlgan}
R.~Noor, A.~Wahid, S.~U. Bazai, A.~Khan, M.~Fang, M.~Syam, U.~A. Bhatti, Y.~Y.
  Ghadi, Dlgan: Undersampled mri reconstruction using deep learning based
  generative adversarial network, Biomedical Signal Processing and Control 93
  (2024) 106218.

\bibitem{li2024progressive}
B.~Li, Z.~Wang, Z.~Yang, W.~Xia, Y.~Zhang, Progressive dual-domain-transfer
  cyclegan for unsupervised mri reconstruction, Neurocomputing 563 (2024)
  126934.

\bibitem{zhou2021efficient}
W.~Zhou, H.~Du, W.~Mei, L.~Fang, Efficient structurally-strengthened generative
  adversarial network for mri reconstruction, Neurocomputing 422 (2021) 51--61.

\bibitem{gungor2023adaptive}
A.~G{\"u}ng{\"o}r, S.~U. Dar, {\c{S}}.~{\"O}zt{\"u}rk, Y.~Korkmaz, H.~A. Bedel,
  G.~Elmas, M.~Ozbey, T.~{\c{C}}ukur, Adaptive diffusion priors for accelerated
  mri reconstruction, Medical image analysis 88 (2023) 102872.

\bibitem{cao2024high}
C.~Cao, Z.-X. Cui, Y.~Wang, S.~Liu, T.~Chen, H.~Zheng, D.~Liang, Y.~Zhu,
  High-frequency space diffusion model for accelerated mri, IEEE Transactions
  on Medical Imaging (2024).

\bibitem{wu2023wavelet}
W.~Wu, Y.~Wang, Q.~Liu, G.~Wang, J.~Zhang, Wavelet-improved score-based
  generative model for medical imaging, IEEE transactions on medical imaging
  (2023).

\bibitem{ahmad2020plug}
R.~Ahmad, C.~A. Bouman, G.~T. Buzzard, S.~Chan, S.~Liu, E.~T. Reehorst,
  P.~Schniter, Plug-and-play methods for magnetic resonance imaging: Using
  denoisers for image recovery, IEEE signal processing magazine 37~(1) (2020)
  105--116.

\bibitem{rasti2023plug}
A.~Rasti-Meymandi, A.~Ghaffari, E.~Fatemizadeh, Plug and play augmented hqs:
  Convergence analysis and its application in mri reconstruction,
  Neurocomputing 518 (2023) 1--14.

\bibitem{kamilov2023plug}
U.~S. Kamilov, C.~A. Bouman, G.~T. Buzzard, B.~Wohlberg, Plug-and-play methods
  for integrating physical and learned models in computational imaging: Theory,
  algorithms, and applications, IEEE Signal Processing Magazine 40~(1) (2023)
  85--97.

\bibitem{wang2021deep}
S.~Wang, T.~Xiao, Q.~Liu, H.~Zheng, Deep learning for fast mr imaging: A review
  for learning reconstruction from incomplete k-space data, Biomedical Signal
  Processing and Control 68 (2021) 102579.

\bibitem{sun2016deep}
J.~Sun, H.~Li, Z.~Xu, et~al., Deep admm-net for compressive sensing mri,
  Advances in neural information processing systems 29 (2016).

\bibitem{zhang2020deep}
K.~Zhang, L.~V. Gool, R.~Timofte, Deep unfolding network for image
  super-resolution, in: Proceedings of the IEEE/CVF conference on computer
  vision and pattern recognition, 2020, pp. 3217--3226.

\bibitem{mou2022deep}
C.~Mou, Q.~Wang, J.~Zhang, Deep generalized unfolding networks for image
  restoration, in: Proceedings of the IEEE/CVF conference on computer vision
  and pattern recognition, 2022, pp. 17399--17410.

\bibitem{ayad2024qn}
I.~Ayad, N.~Larue, M.~K. Nguyen, Qn-mixer: A quasi-newton mlp-mixer model for
  sparse-view ct reconstruction, in: Proceedings of the IEEE/CVF Conference on
  Computer Vision and Pattern Recognition, 2024, pp. 25317--25326.

\bibitem{sriram2020end}
A.~Sriram, J.~Zbontar, T.~Murrell, A.~Defazio, C.~L. Zitnick, N.~Yakubova,
  F.~Knoll, P.~Johnson, End-to-end variational networks for accelerated mri
  reconstruction, in: Medical Image Computing and Computer Assisted
  Intervention--MICCAI 2020: 23rd International Conference, Lima, Peru, October
  4--8, 2020, Proceedings, Part II 23, Springer, 2020, pp. 64--73.

\bibitem{heckel2024deep}
R.~Heckel, M.~Jacob, A.~Chaudhari, O.~Perlman, E.~Shimron, Deep learning for
  accelerated and robust mri reconstruction: a review, arXiv preprint
  arXiv:2404.15692 (2024).

\bibitem{zhang2023camp}
L.~Zhang, X.~Li, W.~Chen, Camp-net: Context-aware multi-prior network for
  accelerated mri reconstruction, arXiv preprint arXiv:2306.11238 (2023).

\bibitem{wang2022one}
Z.~Wang, C.~Qian, D.~Guo, H.~Sun, R.~Li, B.~Zhao, X.~Qu, One-dimensional deep
  low-rank and sparse network for accelerated mri, IEEE Transactions on Medical
  Imaging 42~(1) (2022) 79--90.

\bibitem{huang2021deep}
W.~Huang, Z.~Ke, Z.-X. Cui, J.~Cheng, Z.~Qiu, S.~Jia, L.~Ying, Y.~Zhu,
  D.~Liang, Deep low-rank plus sparse network for dynamic mr imaging, Medical
  Image Analysis 73 (2021) 102190.

\bibitem{xu2024attention}
S.~Xu, K.~Hammernik, A.~Lingg, J.~Kuebler, P.~Krumm, D.~Rueckert, S.~Gatidis,
  T.~Kuestner, Attention incorporated network for sharing low-rank, image and
  k-space information during mr image reconstruction to achieve single
  breath-hold cardiac cine imaging, arXiv preprint arXiv:2407.03034 (2024).

\bibitem{duan2019vs}
J.~Duan, J.~Schlemper, C.~Qin, C.~Ouyang, W.~Bai, C.~Biffi, G.~Bello,
  B.~Statton, D.~P. O’regan, D.~Rueckert, Vs-net: Variable splitting network
  for accelerated parallel mri reconstruction, in: Medical Image Computing and
  Computer Assisted Intervention--MICCAI 2019: 22nd International Conference,
  Shenzhen, China, October 13--17, 2019, Proceedings, Part IV 22, Springer,
  2019, pp. 713--722.

\bibitem{zhang2018ista}
J.~Zhang, B.~Ghanem, Ista-net: Interpretable optimization-inspired deep network
  for image compressive sensing, in: Proceedings of the IEEE conference on
  computer vision and pattern recognition, 2018, pp. 1828--1837.

\bibitem{hammernik2018learning}
K.~Hammernik, T.~Klatzer, E.~Kobler, M.~P. Recht, D.~K. Sodickson, T.~Pock,
  F.~Knoll, Learning a variational network for reconstruction of accelerated
  mri data, Magnetic resonance in medicine 79~(6) (2018) 3055--3071.

\bibitem{yiasemis2022recurrent}
G.~Yiasemis, J.-J. Sonke, C.~S{\'a}nchez, J.~Teuwen, Recurrent variational
  network: a deep learning inverse problem solver applied to the task of
  accelerated mri reconstruction, in: Proceedings of the IEEE/CVF Conference on
  Computer Vision and Pattern Recognition, 2022, pp. 732--741.

\bibitem{li2024gates}
T.~Li, Q.~Yan, Q.~Zou, Q.~Dai, Gates-controlled deep unfolding network for
  image compressed sensing, IEEE Transactions on Computational Imaging (2024).

\bibitem{yu2019review}
Y.~Yu, X.~Si, C.~Hu, J.~Zhang, A review of recurrent neural networks: Lstm
  cells and network architectures, Neural computation 31~(7) (2019) 1235--1270.

\bibitem{shi2015convolutional}
X.~Shi, Z.~Chen, H.~Wang, D.-Y. Yeung, W.-K. Wong, W.-c. Woo, Convolutional
  lstm network: A machine learning approach for precipitation nowcasting,
  Advances in neural information processing systems 28 (2015).

\bibitem{huang2022swin}
J.~Huang, Y.~Fang, Y.~Wu, H.~Wu, Z.~Gao, Y.~Li, J.~Del~Ser, J.~Xia, G.~Yang,
  Swin transformer for fast mri, Neurocomputing 493 (2022) 281--304.

\bibitem{fabian2022humus}
Z.~Fabian, B.~Tinaz, M.~Soltanolkotabi, Humus-net: Hybrid unrolled multi-scale
  network architecture for accelerated mri reconstruction, Advances in Neural
  Information Processing Systems 35 (2022) 25306--25319.

\bibitem{sun2020dual}
L.~Sun, Y.~Wu, B.~Shu, X.~Ding, C.~Cai, Y.~Huang, J.~Paisley, A dual-domain
  deep lattice network for rapid mri reconstruction, Neurocomputing 397 (2020)
  94--107.

\bibitem{liu2023diik}
Y.~Liu, Y.~Pang, X.~Liu, Y.~Liu, J.~Nie, Diik-net: A full-resolution
  cross-domain deep interaction convolutional neural network for mr image
  reconstruction, Neurocomputing 517 (2023) 213--222.

\bibitem{hong2024complex}
T.~Hong, L.~Hernandez-Garcia, J.~A. Fessler, A complex quasi-newton proximal
  method for image reconstruction in compressed sensing mri, IEEE Transactions
  on Computational Imaging (2024).

\bibitem{yiasemis2023vsharp}
G.~Yiasemis, N.~Moriakov, J.-J. Sonke, J.~Teuwen, vsharp: variable splitting
  half-quadratic admm algorithm for reconstruction of inverse-problems, arXiv
  preprint arXiv:2309.09954 (2023).

\bibitem{jiang2023ga}
J.~Jiang, J.~Chen, H.~Xu, Y.~Feng, J.~Zheng, Ga-hqs: Mri reconstruction via a
  generically accelerated unfolding approach, in: 2023 IEEE International
  Conference on Multimedia and Expo (ICME), IEEE, 2023, pp. 186--191.

\bibitem{wang2024progressive}
C.~Wang, L.~Guo, Y.~Wang, H.~Cheng, Y.~Yu, B.~Wen, Progressive
  divide-and-conquer via subsampling decomposition for accelerated mri, in:
  Proceedings of the IEEE/CVF Conference on Computer Vision and Pattern
  Recognition, 2024, pp. 25128--25137.

\bibitem{ehrhardt2024learning}
M.~J. Ehrhardt, P.~Fahy, M.~Golbabaee, Learning preconditioners for inverse
  problems, arXiv preprint arXiv:2406.00260 (2024).

\bibitem{zhang2024t2lr}
Y.~Zhang, P.~Li, Y.~Hu, T2lr-net: An unrolling network learning transformed
  tensor low-rank prior for dynamic mr image reconstruction, Computers in
  Biology and Medicine 170 (2024) 108034.

\bibitem{zhang2020image}
X.~Zhang, D.~Guo, Y.~Huang, Y.~Chen, L.~Wang, F.~Huang, Q.~Xu, X.~Qu, Image
  reconstruction with low-rankness and self-consistency of k-space data in
  parallel mri, Medical Image Analysis 63 (2020) 101687.

\bibitem{2004Learning}
N.~Srebro, Learning with matrix factorizations, massachusetts institute of
  technology (2004).

\bibitem{jin2016general}
K.~H. Jin, D.~Lee, J.~C. Ye, A general framework for compressed sensing and
  parallel mri using annihilating filter based low-rank hankel matrix, IEEE
  Transactions on Computational Imaging 2~(4) (2016) 480--495.

\bibitem{jacob2020structured}
M.~Jacob, M.~P. Mani, J.~C. Ye, Structured low-rank algorithms: Theory,
  magnetic resonance applications, and links to machine learning, IEEE signal
  processing magazine 37~(1) (2020) 54--68.

\bibitem{pramanik2020deep}
A.~Pramanik, H.~K. Aggarwal, M.~Jacob, Deep generalization of structured
  low-rank algorithms (deep-slr), IEEE transactions on medical imaging 39~(12)
  (2020) 4186--4197.

\bibitem{zhang2022accelerated}
X.~Zhang, H.~Lu, D.~Guo, Z.~Lai, H.~Ye, X.~Peng, B.~Zhao, X.~Qu, Accelerated
  mri reconstruction with separable and enhanced low-rank hankel
  regularization, IEEE Transactions on Medical Imaging 41~(9) (2022)
  2486--2498.

\bibitem{ke2021learned}
Z.~Ke, W.~Huang, Z.-X. Cui, J.~Cheng, S.~Jia, H.~Wang, X.~Liu, H.~Zheng,
  L.~Ying, Y.~Zhu, et~al., Learned low-rank priors in dynamic mr imaging, IEEE
  Transactions on Medical Imaging 40~(12) (2021) 3698--3710.

\bibitem{jun2021joint}
Y.~Jun, H.~Shin, T.~Eo, D.~Hwang, Joint deep model-based mr image and coil
  sensitivity reconstruction network (joint-icnet) for fast mri, in:
  Proceedings of the IEEE/CVF Conference on Computer Vision and Pattern
  Recognition, 2021, pp. 5270--5279.

\bibitem{mohan2012iterative}
K.~Mohan, M.~Fazel, Iterative reweighted algorithms for matrix rank
  minimization, The Journal of Machine Learning Research 13~(1) (2012)
  3441--3473.

\bibitem{liu2016projected}
Y.~Liu, Z.~Zhan, J.-F. Cai, D.~Guo, Z.~Chen, X.~Qu, Projected iterative
  soft-thresholding algorithm for tight frames in compressed sensing magnetic
  resonance imaging, IEEE transactions on medical imaging 35~(9) (2016)
  2130--2140.

\bibitem{zhang2021guaranteed}
X.~Zhang, H.~Lu, D.~Guo, L.~Bao, F.~Huang, Q.~Xu, X.~Qu, A guaranteed
  convergence analysis for the projected fast iterative soft-thresholding
  algorithm in parallel mri, Medical Image Analysis 69 (2021) 101987.

\bibitem{fang2024hfgn}
F.~Fang, L.~Hu, J.~Liu, Q.~Yi, T.~Zeng, G.~Zhang, Hfgn: High-frequency residual
  feature guided network for fast mri reconstruction, Pattern Recognition 156
  (2024) 110801.

\bibitem{zhao2015loss}
H.~Zhao, O.~Gallo, I.~Frosio, J.~Kautz, Loss functions for neural networks for
  image processing, arXiv preprint arXiv:1511.08861 (2015).

\bibitem{zbontar2018fastmri}
J.~Zbontar, F.~Knoll, A.~Sriram, T.~Murrell, Z.~Huang, M.~J. Muckley,
  A.~Defazio, R.~Stern, P.~Johnson, M.~Bruno, et~al., fastmri: An open dataset
  and benchmarks for accelerated mri, arXiv preprint arXiv:1811.08839 (2018).

\bibitem{aggarwal2018modl}
H.~K. Aggarwal, M.~P. Mani, M.~Jacob, Modl: Model-based deep learning
  architecture for inverse problems, IEEE transactions on medical imaging
  38~(2) (2018) 394--405.

\bibitem{uecker2014espirit}
M.~Uecker, P.~Lai, M.~J. Murphy, P.~Virtue, M.~Elad, J.~M. Pauly, S.~S.
  Vasanawala, M.~Lustig, Espirit—an eigenvalue approach to autocalibrating
  parallel mri: where sense meets grappa, Magnetic resonance in medicine 71~(3)
  (2014) 990--1001.

\bibitem{ong2019sigpy}
F.~Ong, M.~Lustig, Sigpy: a python package for high performance iterative
  reconstruction, in: Proceedings of the ISMRM 27th Annual Meeting, Montreal,
  Quebec, Canada, Vol. 4819, 2019.

\bibitem{ma2008efficient}
S.~Ma, W.~Yin, Y.~Zhang, A.~Chakraborty, An efficient algorithm for compressed
  mr imaging using total variation and wavelets, in: 2008 IEEE Conference on
  Computer Vision and Pattern Recognition, IEEE, 2008, pp. 1--8.

\bibitem{schlemper2017deep}
J.~Schlemper, J.~Caballero, J.~V. Hajnal, A.~Price, D.~Rueckert, A deep cascade
  of convolutional neural networks for mr image reconstruction, in: Information
  Processing in Medical Imaging: 25th International Conference, IPMI 2017,
  Boone, NC, USA, June 25-30, 2017, Proceedings 25, Springer, 2017, pp.
  647--658.

\bibitem{qiao2023medl}
X.~Qiao, Y.~Huang, W.~Li, Medl-net: A model-based neural network for mri
  reconstruction with enhanced deep learned regularizers, Magnetic Resonance in
  Medicine 89~(5) (2023) 2062--2075.

\bibitem{mokry2023improving}
O.~Mokr{\`y}, J.~Vitou{\v{s}}, P.~Rajmic, R.~Ji{\v{r}}{\'\i}k, Improving
  dce-mri through unfolded low-rank+ sparse optimisation, arXiv preprint
  arXiv:2312.07222 (2023).

\bibitem{lin2018efficient}
C.~Y. Lin, J.~A. Fessler, Efficient dynamic parallel mri reconstruction for the
  low-rank plus sparse model, IEEE transactions on computational imaging 5~(1)
  (2018) 17--26.

\bibitem{dong2020low}
Z.~Dong, X.~Sun, F.~Xu, W.~Liu, A low-rank and sparse decomposition-based
  method of improving the accuracy of sub-pixel grayscale centroid extraction
  for spot images, IEEE Sensors Journal 20~(11) (2020) 5845--5854.

\bibitem{chen2024thick}
Y.~Chen, M.~Chen, W.~He, J.~Zeng, M.~Huang, Y.-B. Zheng, Thick cloud removal in
  multitemporal remote sensing images via low-rank regularized self-supervised
  network, IEEE Transactions on Geoscience and Remote Sensing (2024).

\bibitem{chen2023combining}
Y.~Chen, X.~Gui, J.~Zeng, X.-L. Zhao, W.~He, Combining low-rank and deep
  plug-and-play priors for snapshot compressive imaging, IEEE Transactions on
  Neural Networks and Learning Systems (2023).

\bibitem{yuan2020sara}
Z.~Yuan, M.~Jiang, Y.~Wang, B.~Wei, Y.~Li, P.~Wang, W.~Menpes-Smith, Z.~Niu,
  G.~Yang, Sara-gan: Self-attention and relative average discriminator based
  generative adversarial networks for fast compressed sensing mri
  reconstruction, Frontiers in Neuroinformatics 14 (2020) 611666.

\bibitem{wu2019self}
Y.~Wu, Y.~Ma, J.~Liu, J.~Du, L.~Xing, Self-attention convolutional neural
  network for improved mr image reconstruction, Information sciences 490 (2019)
  317--328.

\bibitem{zhou2021spatial}
W.~Zhou, H.~Du, W.~Mei, L.~Fang, Spatial orthogonal attention generative
  adversarial network for mri reconstruction, Medical Physics 48~(2) (2021)
  627--639.

\bibitem{huang2019mri}
Q.~Huang, D.~Yang, P.~Wu, H.~Qu, J.~Yi, D.~Metaxas, Mri reconstruction via
  cascaded channel-wise attention network, in: 2019 IEEE 16th International
  Symposium on Biomedical Imaging (ISBI 2019), IEEE, 2019, pp. 1622--1626.

\bibitem{liu2022coil}
J.~Liu, C.~Qin, M.~Yaghoobi, Coil-agnostic attention-based network for parallel
  mri reconstruction, in: Proceedings of the Asian Conference on Computer
  Vision, 2022, pp. 2866--2883.

\bibitem{liu2022dual}
X.~Liu, Y.~Pang, R.~Jin, Y.~Liu, Z.~Wang, Dual-domain reconstruction network
  with v-net and k-net for fast mri, Magnetic Resonance in Medicine 88~(6)
  (2022) 2694--2708.

\bibitem{zhou2023rnlfnet}
L.~Zhou, M.~Zhu, D.~Xiong, L.~Ouyang, Y.~Ouyang, Z.~Chen, X.~Zhang, Rnlfnet:
  Residual non-local fourier network for undersampled mri reconstruction,
  Biomedical Signal Processing and Control 83 (2023) 104632.

\bibitem{das2023recurrent}
S.~Das, A.~Tariq, T.~Santos, S.~S. Kantareddy, I.~Banerjee, Recurrent neural
  networks (rnns): architectures, training tricks, and introduction to
  influential research, Machine Learning for Brain Disorders (2023) 117--138.

\bibitem{chen2022pyramid}
E.~Z. Chen, P.~Wang, X.~Chen, T.~Chen, S.~Sun, Pyramid convolutional rnn for
  mri image reconstruction, IEEE Transactions on Medical Imaging 41~(8) (2022)
  2033--2047.

\bibitem{song2021memory}
J.~Song, B.~Chen, J.~Zhang, Memory-augmented deep unfolding network for
  compressive sensing, in: Proceedings of the 29th ACM international conference
  on multimedia, 2021, pp. 4249--4258.

\end{thebibliography}

\end{document}